\documentclass[journal,onecolumn,draftcls]{IEEEtran}

\newcommand{\StyleSingleDraft}[0]{\StyleDoublefalse\StyleSingletrue}
\newif\ifStyleDouble\StyleDoubletrue
\newif\ifStyleSingle\StyleSingletrue

\StyleSingleDraft

\usepackage{calc}
\newcommand{\figwidth}{\minof{\columnwidth}{4.0in}}

\usepackage{cite}
\usepackage{graphicx}
\graphicspath{{./figures/}}

\ifCLASSINFOpdf
  \usepackage{graphicx}
  \graphicspath{{./figures/}}
\else
\fi

\usepackage[cmex10]{amsmath}
\usepackage{multirow}
\usepackage{booktabs}

\begin{document}
\title{Spectrum Sharing Between A Surveillance Radar and Secondary Wi-Fi Networks}

\author{Farzad~Hessar,~\IEEEmembership{Student Member,~IEEE,}
				Sumit~Roy,~\IEEEmembership{Fellow,~IEEE,}
\\ Electrical Engineering Department, University of Washington
\\ \{farzad, sroy\}@u.washington.edu
\thanks{Contact Author: farzad@u.washington.edu}
\thanks{This work was supported in part by AFRL via Task CRFR-009-02-01 and NSF AST 1443923}\\
}

\maketitle
\thispagestyle{empty}

\begin{abstract}

Co-existence between unlicensed networks that share spectrum spatio-temporally with terrestrial (e.g. Air Traffic Control) and shipborne radars\footnote{In the US, airport radars are allocated 2700-2900 MHz, and 3100-3650 MHz for military radar operations for national defense.} in $3$ GHz band is attracting significant interest. Similar to every primary-secondary coexistence scenario, interference from unlicensed devices to a primary receiver must be within acceptable bounds. In this work, we formulate the spectrum sharing problem between a pulsed, search radar (primary) and 802.11 WLAN as the secondary. We compute the protection region for such a search radar for a) a single secondary user (initially) as well as b) a random spatial distribution of  multiple secondary users.  Furthermore, we also analyze the interference {\em to} the WiFi devices from the radar's transmissions to estimate the impact on achievable WLAN throughput as a function of distance to the primary radar.

\end{abstract}


\IEEEpeerreviewmaketitle
\section{Introduction}
Wireless data traffic has been increasing exponentially over the last decade, resulting from the boom in multimedia applications running on high-end client devices such as smart phones, tablets \cite{ch5:ele:Cisco}. Various solutions are suggested for expanding capacity of wireless networks, from higher spectral efficiency to smaller cell sizes; utilizing additional spectrum is always a major element of the solution. The scarcity of available new RF spectrum and technological limitations for usage of higher frequency bands (above 60 GHz) has led to a renewed emphasis on more efficient use of existing spectrum. This has motivated spectrum regulatory bodies such as FCC (US) and Ofcom (UK) to promulgate {\em dynamic access} rules by smart secondary devices. This allows {\em cognitive secondary users} to detect locally unused {\em white spaces} and use them for a period, subject to agreed upon rules of primary protection \cite{ch5:Std:FCC12,ch5:paper:farzad14}.

In this work, we focus on spectrum sharing between primary radar systems and secondary 802.11 WLAN networks - a topic on which little work exists beyond the studies in \cite{ch5:paper:Tercero13,ch5:darpa:ssparc,ch5:paper:peha2012,ch5:thesis:Peha2012,ch5:conf:peha-vtc13,ch5:conf:Paisana14,ch5:paper:Tercero11,ch5:conf:Shajaiah14,ch5:conf:rahman11,ch5:paper:deng13}. The re-emergence of interest in this topic is based in part on large amount of licensed spectrum allocated to radar operations in the U.S. - over 1700 MHz  in 225 MHz to 3.7 GHz band, are set aside for radar and radio-navigation \cite{ch5:ppt:ntia-beachfront} and the widespread deployment of 802.11 WLAN networks. Given that terrestrial radar locations are fixed and have predictable operational patterns, it is possible to model their behavior and utilize it for a database-driven coexistence solution,  akin to the architecture espoused by the FCC for TV white spaces \cite{ch5:ele:FccSAS}.

Database-driven spectrum sharing uses a geo-location database that determines available spectrum for a secondary user (SU) requesting access based on their location. This coexistence mechanism is currently mandated by the FCC for operation in UHF TV bands; its main impact was to remove the burden of spectrum sensing from secondary devices thereby  simplifying receiver design for clients and also avoiding other challenges in distributed spectrum sensing such as the well-known hidden terminal problem. By rules of cognitive access, overlay secondary users are prohibited from re-using a primary operating channel within an area defined as the {\em protection region}. The geo-location database has access to relevant information of primary users such as location, transmit power, interference tolerance, etc. that it utilizes to estimate this protection region to enable any secondary transmitter to meet the interference protection conditions.

The actual implementation of any incumbent protection rule depends strongly on the usage scenario, i.e., features of the primary and secondary systems and the consequent co-existence requirements. In this paper, we consider a  {\em rotating search radar} as the licensed transmitter and WiFi networks as unlicensed devices. First, we review the known design equations that represent performance characteristics of a typical search radar for the purely noise limited case in terms of the desired probability of detection $P_D$ and false alarm $P_{FA}$. This determines the minimum SNR requirements at the boundary of the radar operating range and sets the baseline for comparison with any spectrum sharing regime.

In order to permit overlay transmission by secondaries, we need to define the rules for co-existence. A recent program suggests drop of 5\% in $P_D$ for fixed $P_{FA}$ \cite{ch5:darpa:ssparc} at the edge of radar operating range as being acceptable; this defines the protection regime for the primary receiver (from secondary interference). However, the fundamental objective of any WS type spectrum sharing scenario is to {\em promote} secondary usage subject to the primary protection constraints; we thus also analyze the effect of (high power) radar pulse sequences on the throughput of WiFi network. Any successful spectrum sharing system must balance the rights of the incumbent (primary protection) with encouraging new services, and we hope that our work fundamentally highlights the inherent trade-offs in this design space.

\subsection{Related Works}

There is growing interest in radar spectrum sharing from both regulators and researchers \cite{ch5:paper:Tercero13,ch5:darpa:ssparc,ch5:paper:peha2012,ch5:thesis:Peha2012,ch5:conf:peha-vtc13,ch5:conf:Paisana14,ch5:paper:Tercero11,ch5:conf:Shajaiah14,ch5:conf:rahman11,ch5:paper:deng13,ch5:ofcom:2009,ch5:ntia:14499,ch5:itu:1464-1,ch5:itu:1638,ch5:ntia:13490,ch5:ntia:5GhzRadar,ch5:paper:Griffiths15,ch5:ntia:rf-radar-interf}. SSPARC program from DARPA\cite{ch5:darpa:ssparc} is a good example that seeks to support two types of sharing: a) Military/military sharing between military radars and military communication systems to increase capabilities of both and b) Military/commercial sharing between military radars and commercial communication systems to preserve radar capabilities while meeting the need for increased capacity of commercial networks.

In  \cite{ch5:paper:peha2012,ch5:thesis:Peha2012,ch5:conf:peha-vtc13}, the authors study coexistence between radar and a cellular base station. The co-existence strategy espoused is variable secondary
transmit power {\em assuming a maximum tolerable interference at the radar}. Further, the authors consider only one sharing scenario in which SU is perfectly synchronized with radar rotation (a very impractical assumption). The limitations of this analysis is thus apparent - it does not explore at any depth, how the radar's interference tolerance is determined based on the system parameters and geometry considerations. Similarly in \cite{ch5:paper:Tercero13}, temporal variations of radar antenna's main lobe is exploited to support more white space users when their location is not within the main lobe. In \cite{ch5:conf:Paisana14}, spectrum sensing is combined with database approach to create a hybrid spectrum sharing technique. The authors in \cite{ch5:conf:rahman11} study the potential for secondary LTE usage in 2.7-2.9 GHz radar bands for different scenarios such as home eNodeB (HeNB) at street levels or in high-rise buildings, macro LTE transmitters and so on. A fixed Interference-to-Noise ratio (INR) of -10 dB  is specified for sharing without any discussion on radar performance. The analysis does not consider radar rotation and mostly focused on single-user sharing with radar. While the case of multiple SU is considered, it is done so under an unrealistic assumption that all users are at the same distance from radar. Finally, spectrum sharing between a MIMO radar and a wireless communication system is analyzed in \cite{ch5:paper:deng13}. Their interference mitigation approach is shown to eliminate wireless interferences from main/side lobe while maintaining target detection performance.

 Some U.S. DoD studies for co-existence with radars operating in the 2700-2900 MHz and 5250-5850 MHz bands  \cite{ch5:itu:1464-1,ch5:itu:1638} have also been conducted. Protection criteria against external interference is determined through experimental measurements by injecting three types of unwanted communications waveform emissions - continuous wave, CDMA-QPSK, and TDMA-QPSK. In \cite{ch5:ntia:13490}, the authors evaluate interference from broadband communication transmitters such as WiMax to  WSR-88D next-generation weather radar (2700-2900 MHz). A computation model for calculating aggregate interference from radio local area networks to 5-GHz radar systems is provided in \cite{ch5:ntia:5GhzRadar}. The analysis methodology is based on using point to point path loss models between radio networks and radar as well as other link parameters such as antenna gains and frequency-dependent rejection \cite{ch5:ofcom:2009, ch5:ntia:14499}.

Our major contribution in this work is a {\em complete characterization} of Radar - WiFi coexistence as a function of all relevant system parameters and design constraints/objectives.
First, the maximum tolerable interference {\em from} WiFi networks {\em to} radar is estimated for both a
a) single WiFi network and b) a (random) spatial distribution of multiple WiFi networks.
Depending on how much information about radar is available to secondary (WiFi) networks, various sharing scenarios are considered, resulting in different protection distances. Second,the (time-varying)  interference {\em from}  radar {\em to} WiFi networks is modeled and achievable secondary link throughput is estimated.

The rest of this paper is organized as follows. In section II, baseline performance for a noise limited radar is formulated. Section III considers coexistence between radar and a single SU. Multiple SU with spatial distribution is discussed in IV. In section V, interference from radar to SU is studied as a limiting factor to available white space capacity. Numerical results are provided in VI and finally VII concludes the paper.

\section{Search Radar: Noise Limited Operation\cite{ch5:book:radar}}

We first review operational characteristics of a typical search radar in the noise limited regime with no external source of interference. For a radar transmitting a pulse train $x(t) = \sum_{n}\sqrt{P_T}s(t-\frac{n}{f_R})$ with instantaneous power $P_T$ and pulse repetition frequency of $f_R$, the power of reflected signal from the target at the radar receiver, assuming free space propagation is given by the well-know {\em Radar Equation}, i.e.,
\begin{equation}
P_R = \frac{P_T G^2 \lambda^2}{(4\pi)^3 d^4} \sigma
\label{ch5:eq:radar-eq}
\end{equation}
where $G$ is the radar's antenna gain (relative to isotropic antenna) on both transmit and receive, $\lambda$ is the wavelength and $d$ the distance from source to the target of interest, and $\sigma$ represents the target's {\em radar cross section}.

For a single received pulse, the signal-to-noise ratio (SNR) at the receiver input is calculated as
\begin{equation}
\mbox{SNR}_p = \frac{P_T G^2 \lambda^2}{(4\pi)^3 d^4 N_0 f_{BW}} \sigma
\label{ch5:eq:pulse-snr}
\end{equation}
with $f_{BW}$ representing the pulse bandwidth and $N_0$ being the one-sided noise spectral density.
\begin{equation}
N_0 = F K T_E
\label{ch5:eq:N0}
\end{equation}
where $F$ is the receiver noise figure and $T_E$ is the ambient temperature. Radar detection typically operates based on processing of multiple pulses received from the target. For a {\em coherent} radar receiver that uses $M$ pulses, the energy of the pulses are integrated such that the resulting SNR at the detector input is increased by a factor of $M$, i.e.,
\begin{equation}
\mbox{SNR}_{\mbox{eff}} = M \frac{P_T G^2 \lambda^2}{(4\pi)^3 d^4 N_0 f_{BW}} \sigma
\label{ch5:eq:total-snr}
\end{equation}
where $M = T_I \; f_R$, product of illumination time $T_I$ and pulse repetition frequency $f_R$. The target illumination time $T_I$ depends on radar scan rate as well as antenna pattern. Let $\theta_V$ and $\theta_H$ (in radian) denote the vertical and horizontal antenna beam width, respectively, then the antenna gain can be approximated as
\begin{align}
G \approx \frac{4\pi}{\theta_H \theta_V}\rho_A
\label{ch5:eq:antenna-gain}
\end{align}
where $\rho_A$ is the antenna efficiency, i.e., the radar antenna is concentrating an otherwise uniformly distributed power into an area of $\theta_V\theta_H$ with efficiency of $\rho_A$ where the latter is typically around $0.5$. If radar is scanning over an area of $\Omega$ (steradians), within a scan time of $T_S$, then illumination time is determined as:
\begin{equation}
T_I \approx T_S\frac{\theta_H \theta_V}{\Omega} \approx T_S\frac{4\pi\rho_A}{\Omega G}
\label{ch5:eq:illum-time}
\end{equation}
For a radar that searches the entire azimuth/elevation plane, $\Omega = 4\pi$.

Using (\ref{ch5:eq:total-snr})-(\ref{ch5:eq:illum-time}) yields
\begin{equation}
\mbox{SNR}_{\mbox{eff}} = \frac{T_S}{\Omega} \frac{P_T G \lambda^2 f_R}{(4\pi)^2 d^4 N_0 f_{BW} L} \sigma
\label{ch5:eq:SNR}
\end{equation}
Here, antenna efficiency $\rho_A$ is replaced by $L$ that represents total losses in the system, including antenna efficiency, transmission lines mismatch, perfect coherence in pulse detector, etc.

\subsection{Minimum Required SNR}
Radar detection performance is defined in terms of two probabilities, detection $P_D$ and false alarm $P_{FA}$, which in turn depend on SNR at the detector input. The latter is determined by the pulse integration method that is utilized by the receiver, namely {\em coherent} versus {\em non-coherent}.

 a) \underline{Single pulse, hard detection}: If the received signal at the detector input is
\begin{equation}
e_0(t) = r(t)\cos\left(\omega_c t + \phi(t)\right)
\label{ch5:eq:detector-input}
\end{equation}
then the PDF of the detected envelope for a single pulse is {\em Rician}, i.e.,
\begin{equation}
p(r) = \frac{r}{\beta^2}e^{\frac{-(r^2+A^2)}{2\beta^2}} I_0\left(\frac{rA}{\beta^2} \right)
\label{ch5:eq:detector-input-pd}
\end{equation}
where $A$ is the amplitude of the base band pulse and $\beta=\sqrt{N_0 f_{BW}}$. Therefore, $P_{FA}$ is determined by setting $A=0$ and integrating over 0 to detection threshold $V_T$ as:
\begin{equation}
P_{FA} = e^{\frac{-V_T^2}{2\beta^2}}
\label{ch5:eq:FA}
\end{equation}
A similar general closed-form equation for $P_D$ is complicated. However, for {\em high-SNR} cases, $p(r)$ is well approximated as Gaussian, for which case $P_D$ is given by \cite{ch5:book:radar}
\begin{equation}
P_{D} = \frac{1}{2}\left[ 1 - erf\left( \frac{V_T}{\beta\sqrt{2}} - \sqrt{\mbox{SNR}_p} \right) \right]
\label{ch5:eq:PD}
\end{equation}
The relationship between $P_D, P_{FA}$ and $SNR$ is fairly accurately expressed via the following empirical equation \cite{ch5:book:radar}:
\ifStyleDouble
\begin{align}
\mbox{SNR}_p =& \ln{\left(\frac{0.62}{P_{FA}}\right)} + 0.12\ln{\left(\frac{0.62}{P_{FA}}\right)}\ln{\left(\frac{P_D}{1-P_D}\right)} \nonumber \\
             &+ 1.7\ln{\left(\frac{P_D}{1-P_D}\right)}
\label{ch5:eq:snr-pd-pfa}
\end{align}
\else
\begin{align}
\mbox{SNR}_p = \ln{\left(\frac{0.62}{P_{FA}}\right)} + 0.12\ln{\left(\frac{0.62}{P_{FA}}\right)}\ln{\left(\frac{P_D}{1-P_D}\right)} + 1.7\ln{\left(\frac{P_D}{1-P_D}\right)}
\label{ch5:eq:snr-pd-pfa}
\end{align}
\fi

b) \underline{Coherent Integrator}:
For a coherent receiver integrating $M$ pulses, the SNR-performance relationship is described in (\ref{ch5:eq:snr-pd-pfa}) in which $\mbox{SNR}_p$ should be replaced with the effective SNR at the detector input ($\mbox{SNR}_{\mbox{eff}}$), determined by (\ref{ch5:eq:SNR}). \\

c) \underline{Noncoherent Integrator}:
If radar utilizes a linear (rather than square-law) detector for single pulse and then combines $M$ pulses non-coherently, the required SNR per pulse for desired $P_D, P_{FA}$ is \cite{ch5:paper:approx-pd-pfa-snr}:
\ifStyleDouble
\begin{align}
\mbox{SNR}_{p, dB} &= -5\log_{10}(M) + \left[ 6.2 + \frac{4.54}{\sqrt{M+0.44}} \right]\times \nonumber \\ 
                   &\;\;\;\; \log_{10}(A+0.12AB+1.7B) \nonumber \\
                 A &= \ln{\frac{0.62}{P_{FA}}}, B = \ln{\frac{P_D}{1-P_D}}
\label{ch5:eq:noncoh-integ}
\end{align}
\else
\begin{align}
\mbox{SNR}_{p, dB} &= -5\log_{10}(M) + \left[ 6.2 + \frac{4.54}{\sqrt{M+0.44}} \right] \log_{10}(A+0.12AB+1.7B) \nonumber \\
                 A &= \ln{\frac{0.62}{P_{FA}}}, B = \ln{\frac{P_D}{1-P_D}}
\label{ch5:eq:noncoh-integ}
\end{align}
\fi

Overall, the baseline performance of a noise-limited radar can be evaluated in two ways:
\begin{itemize}
	\item Assuming that maximum operational range of the radar is known, calculate SNR from (\ref{ch5:eq:pulse-snr}) or (\ref{ch5:eq:SNR}) for the maximum distance $d$. Then, using either (\ref{ch5:eq:snr-pd-pfa}) or (\ref{ch5:eq:noncoh-integ}), we can trade-off between $P_D$ and $P_{FA}$.
	\item Assuming that target $P_D$ and $P_{FA}$ is specified, estimate required SNR from (\ref{ch5:eq:snr-pd-pfa}) or (\ref{ch5:eq:noncoh-integ}) and then determine maximum range from (\ref{ch5:eq:pulse-snr}) or (\ref{ch5:eq:SNR}).
\end{itemize}
For our calculations in the following sections, we consider a detector with coherent integration, using effective SNR in (\ref{ch5:eq:SNR}) with (\ref{ch5:eq:snr-pd-pfa}).

\section{Interference Limited Radar - Single Secondary}

In this section, we consider spectrum sharing with a single Wi-Fi user as the secondary device by treating secondary signals as an external interference to radar receiver. Wi-Fi transmissions use OFDM signals, whereby each OFDM symbol is a linear combination of many randomly modulated sub-carriers. Hence, using central limit theorem, each sample of OFDM signal in time-domain is well-approximated as a Gaussian random variable. The matched filter utilized at radar front-end for pulse detection applies another linear transformation on this OFDM signal and results in a Gaussian random variable which is independent of AWGN (thermal noise) at the radar receiver\cite{ch5:paper:Ochiai01}. Therefore, the interference power can be directly added to AWGN noise power, effectively raising the noise floor. Thus system performance is determined  by Signal-to-Interference-plus-Noise ratio (SINR) at radar receiver input. Using (\ref{ch5:eq:SNR}),  this is given by 
\begin{equation}
\mbox{SINR} = \frac{T_S}{\Omega} \frac{P_T G \lambda^2 f_R}{(4\pi)^2 d^4 L \left(N_0f_{BW} + I\right)} \sigma
\label{ch5:eq:SINR}
\end{equation}
where $I$ represents total interference power received from secondary user. The latter depends on various factors: the distance and frequency dependent path loss between secondary source and radar receiver, the azimuth between SU direction and radar's main antenna beam, etc. as below:
\begin{align}
I_{SU\rightarrow Radar} = \frac{P_{SU} G(\alpha_H, \alpha_V)}{L_1(d_{Rd-SU}) \mbox{FDR}(\Delta f)}
\label{ch5:eq:singleIntrf}
\end{align}
where $G(\alpha_H, \alpha_V)$ defines radar's antenna gain in the direction of SU (considering azimuth and elevation), $\mbox{FDR}(\Delta f)$ is {\em frequency dependent rejection} factor that depends on spectral shape of transmitted signal $P(f)$ and receiver receive input filter $H(f)$, i.e.
\begin{align}
\mbox{FDR}(\Delta f) = \frac{\int_0^{\infty}{P(f)df}}{\int_0^{\infty}{P(f)H(f+\Delta f)df}}
\end{align}
 represents the out-of-band emission from the WiFi source into the radar RF receiver front-end as a function of $\Delta f = f_t-f_r$, the difference between interferer and receiver tuned center frequency. For a special case of exact co-channel operation $\Delta f=0$; for a perfectly flat filter response $H(f)=1$, FDR simplifies as the ratio of WiFi to radar bandwidth:
\begin{align}
\mbox{FDR} = \max \left( \frac{\mbox{WiFi BW}}{f_{BW}}, 1 \right)
\end{align}
Our focus in this paper is cases where radar bandwidth is less than WiFi bandwidth, $\mbox{FDR} \ge 1$. 

The minimum required SINR for normal operation of the radar was defined in previous section. Therefore maximum additional interference level $I$ that can be tolerated is determined as:
\begin{align}
\mbox{SINR}_0 &\leq \frac{T_S}{\Omega} \frac{P_T G \lambda^2 f_R \sigma}{(4\pi)^2 d^4 L \left(N_0f_{BW} + I\right)} \nonumber \\
I &\leq \frac{T_S}{\Omega} \frac{P_T G \lambda^2 f_R \sigma}{(4\pi)^2 d^4 L \, \mbox{SINR}_0} - N_0f_{BW} = I_{\max}
\label{ch5:eq:maxI}
\end{align}
Using (\ref{ch5:eq:singleIntrf}) and (\ref{ch5:eq:maxI}), we can calculate the minimum separation distance between radar and SU\footnote{Or equivalently maximum transmission power for a known distance} as:
\begin{align}
d_{Rd-SU} \geq L_{Rd-SU}^{-1}\left( \frac{P_{SU} G(\alpha_H, \alpha_V)}{ \mbox{FDR}(\Delta f) I_{\max}}\right)
\label{ch5:eq:distance}
\end{align}
Where $L_{Rd-SU}(.)$ is the path-loss between radar and SU as a function distance. As this equation suggests, the minimum separation distance depends on the instantaneous antenna gain, $G(\alpha_H, \alpha_V)$.

\subsection{Numerical Results}

For the computations in this section, we use radar parameters from ITU document, Rec. ITU-R M.1464-1, for a typical aeronautical radio-navigation radar in 2.8 GHz band. Table \ref{ch5:tb:rdr-aeronautical-B} in Appendix provides parameters for the so-called type-B radar in \cite{ch5:itu:1464-1}. We choose performance points (of ROC) shown in Table \ref{ch5:tb:rdr-performance} for the radar in noise and interference limited cases.
\begin{table}[ht!]
\caption{Target ROC for Noise/Interference Limited Performance}
\label{ch5:tb:rdr-performance}
\centering
\begin{tabular}{|c|c|c|c|}
\hline
\hline
{\bf Mode } & {\bf $P_D$ } & {\bf $P_{FA}$ } & {SNR/SINR (dB) }\\
\hline
Noise Limited & 0.90 & $10^{-6}$ & 13.14\\
\hline
Interference Limited & 0.85 & $10^{-6}$ & 12.80 \\
\hline
\hline
\end{tabular}
\end{table}
As suggested by SSPARC, a drop of 5\% in performance is permitted to provide an interference margin for the secondary user, which is equivalent to an SNR loss of 0.34 dB.

To calculate max allowable interference level, $I_{\max}$ in (\ref{ch5:eq:maxI}), maximum operational range $d$ and minimum target's radar cross section $\sigma$ are required which is not provided by Table \ref{ch5:tb:rdr-aeronautical-B}. Any variation in values of these two parameters can significantly affect resulting $I_{\max}$. For example, if the SNR of the noise-limited regime in (\ref{ch5:eq:SNR}) is 16.14 dB (3-dB above the required SNR in table \ref{ch5:tb:rdr-performance}) then $I_{\max}$ can be about as high as noise level $N_0f_{BW}$ (INR of $0$ dB), which brings SINR down to $13.14$ dB. However, if we assume that SNR is already at the minimum level, then we only have $0.34$ dB room for the interference which reduces the maximum INR down to $-11$ dB.

The type-B radar, as outlined in \cite{ch5:itu:1464-1}, employs high and low-beam horns in the antenna feed array. The high-beam horn receives returns from high-altitude targets close to the antenna, while the low-beam horn receives returns from low-altitude targets at greater distances. Overall, it is designed for monitoring air traffic in and around airports within a range of 60 Nm (approximately 111 km). A coverage pattern is also provided for a target with 1 $m^2$ radar cross section. Therefore, using $d = 111$ km and $\sigma=1$, the resulting SNR (\ref{ch5:eq:SNR}) will be 30.6 dB that is significantly bigger than required SNR of 13.14 dB. This is unrealistic and does not represent radar's borderline operation. Therefore, in order to remove the effect of $d$ and $\sigma$ in our calculation, we normalize them such that SNR in (\ref{ch5:eq:SNR}) matches with required SNR in (\ref{ch5:eq:snr-pd-pfa})\footnote{By increasing the value of $d$ or decreasing $\sigma$, effective SNR is reduced to match with (\ref{ch5:eq:snr-pd-pfa})}. Based on these normalized parameter values, Table \ref{ch5:tb:Imax} shows the maximum permitted interference level and resulting INR. For noise limited case, no external interference is allowed because the radar's performance is already at the edge.

The results of two administrative tests, performed in \cite{ch5:itu:1464-1} by injecting three types of interfering signals (Continuous Wave, CDMA-QPSK, TDMA-QPSK) to radar's receiver input, have also concluded that and an INR of -10 dB can fully protect radar type B and other aeronautical radionavigation radars operating in the 2700-2900 MHz.
\begin{table}[ht!]
\caption{Maximum Permitted Interference Level}
\label{ch5:tb:Imax}
\centering
\begin{tabular}{|c|c|c|c|}
\hline
\hline
{\bf Mode } & $I_{\max}$ (dBm) & INR (dB) \\
\hline
Noise Limited & $-\infty$ & $-\infty$ \\
\hline
Interference Limited & -122.64 & -10.96 \\
\hline
\hline
\end{tabular}
\end{table}

\subsection{Protection Distance}
The maximum interference level that was calculated in previous section can be used in (\ref{ch5:eq:distance}) to define minimum separation between SU and radar receiver. SU is assumed to be a Wi-Fi AP with following parameters:
\begin{table}[ht!]
\caption{Secondary User Specification}
\label{ch5:tb:SU-param}
\centering
\begin{tabular}{|c|c|}
\hline
\hline
{\bf Parameter } & {\bf Value} \\
\hline
Emission Power (EIRP) $P_{SU}$ & 1 Watt\\
Bandwidth (MHz) & 20.0\\
Antenna Height (m) & 3.0\\
Interference Type & Co-channel, $\Delta f=0$ \\
Antenna Gain (Dipole) & 2.15 dBi \\
Noise Figure (dB) & 8.0 \\
\hline
\hline
\end{tabular}
\end{table}

Since SU and Radar are assumed to be co-channel, the FDR is calculated as the ratio of corresponding bandwidths $\mbox{FDR}=\frac{20 MHz}{653 KHz} = 30.6$. Note that we have used IF 3-dB bandwidth for radar's receiver which is significantly smaller than RF 3-dB bandwidth of 10 MHz, since radar signal detection happens at IF.

\begin{figure}[!t]%
	\centering
	\includegraphics[width=\figwidth]{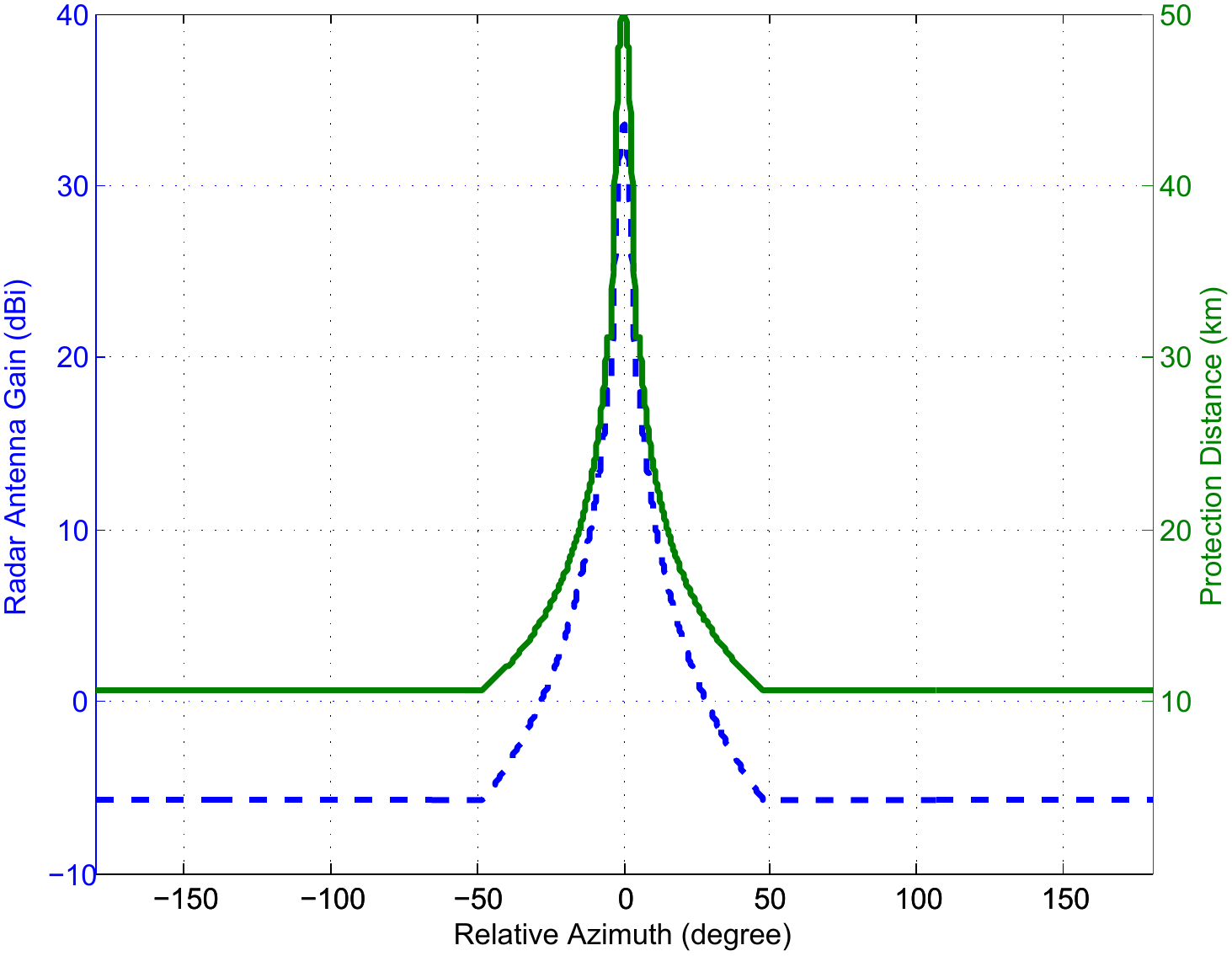}
	\caption{Radar antenna gain as well as Protection distance between SU and Radar v.s. azimuth.}
	\label{ch5:fig:radarGainPrtRgn}
\end{figure}

A statistical antenna gain model is introduced in \cite{ch5:ntia:5GhzRadar} to determine the radar antenna gain in the azimuth and elevation orientations. For high gain values of $22<G_{\max}=33.5<48$ dBi, following piece-wise function is suggested:
\begin{align}
G(\theta) = \left\{
            \begin{array}{lr}
								G_{\max} - 0.0004*10^{G_{\max}/10}\theta^2 & \theta\in[0, \theta_M] \\
								0.75G_{\max} - 7                           & \theta\in[\theta_M, \theta_R] \\
								53 - G_{\max}/2 - 25\log(\theta)           & \theta\in[\theta_R, \theta_B] \\
								11 - G_{\max}/2                            & \theta\in[\theta_B, 180]
            \end{array}
						\right.
\label{ch5:eq:radargain}
\end{align}
where $\theta_M = 50\sqrt{0.25G_{\max}+7}/10^{G_{\max}/20}$, $\theta_R=250/10^{G_{\max}/20}$ and $\theta_B=48$. Figure \ref{ch5:fig:radarGainPrtRgn} shows antenna gain versus azimuth with a main lob of 33.5 dBi. The 3-dB beam width in this pattern is 3.7 degree. Using this antenna pattern and Longley-Rice path loss model\cite{ch5:ITS:4Model} between secondary user and radar's receiver, Figure \ref{ch5:fig:radarGainPrtRgn} also shows protection region as a function of relative azimuth between SU and radar's antenna main beam. {\em It is clear that protection region follows the same pattern as radar's antenna pattern as suggested by (\ref{ch5:eq:distance})}.

\section{Interference Limited Radar - Multiple Secondary Networks}

Equation (\ref{ch5:eq:maxI}) defines a maximum interference level that a radar can tolerate while its performance is in the acceptable range. From radar's point, if multiple secondary users coexist with the radar simultaneously, the accumulated signal power at radar's location must also be bounded by (\ref{ch5:eq:maxI}). In practice, this is the more common scenario due the proliferation of WiFi networks.

\begin{figure}[!t]%
	\centering
	\includegraphics[width=\figwidth]{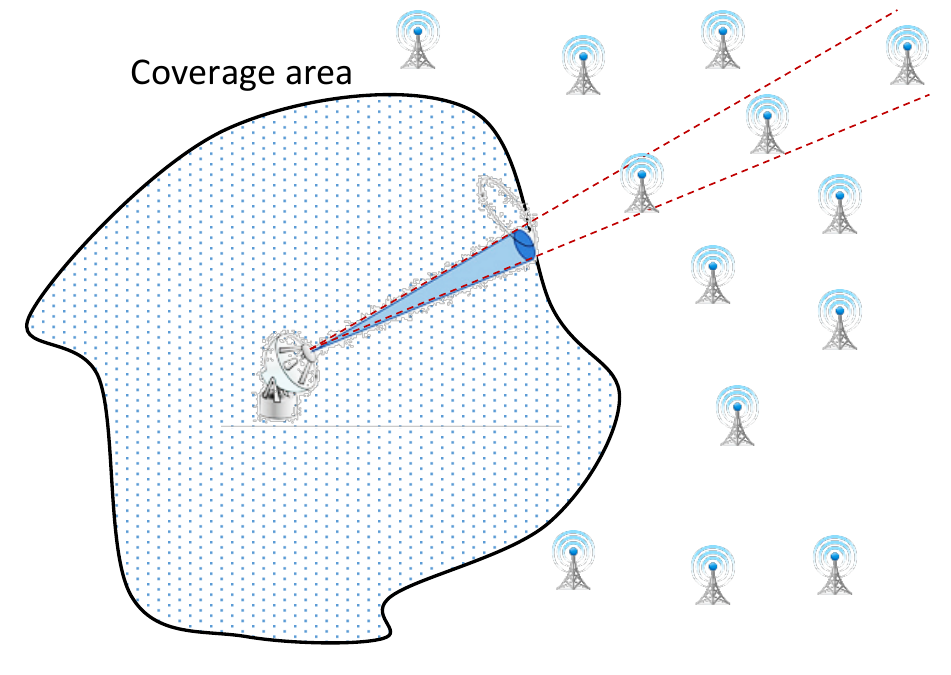}
	\caption{Aggregate interference from multiple WiFi access points to radar receiver.}
	\label{ch5:fig:multiAP}
\end{figure}

The characterization of the aggregate interference from multiple WiFi APs as seen by a radar receiver, is fundamentally determined by the multiple access protocol employed by WiFi nodes. Users within a single WiFi network {\em time-share} the common channel based on CSMA-CA, i.e., the  WiFi DCF protocol prohibits simultaneous multiple user transmissions. However, {\em different} WiFi networks can simultaneously operate in the vicinity of a radar and the aggregate interference across different networks needs to be accounted for, as shown in the scenario in Figure \ref{ch5:fig:multiAP}.

Wi-Fi APs and their associated users are randomly distributed in space which is suitably modeled as a Poisson Point Process (PPP). For a radar located at point $y \in R^d$ and randomly distributed access points at $x\in R^d$, the aggregate interference from secondary users to radar is described as a generalized shot noise process in space \cite{ch5:Haenggi:2009:ILW}:
\begin{align}
I_{aggr}(y) = \sum_{x \in \Phi}{\frac{P_x G(\theta_x)}{\mbox{FDR}(\Delta f)} l(||y - x||)}
\label{ch5:eq:shotnoise}
\end{align}
where $P_x$ is the SU transmit power at location $x$ (an i.i.d random variable) and $||y - x||$ describes the distance between radar and secondary access point. $l(.)$ is the impulse response function that models signal attenuation (inverse of path loss) and $G(.)$ is radar's antenna gain in the direction of interferer, $\theta_x$. Stochastic distribution of $I_{aggr}$, plays an important role in performance analysis for radar. We assume all WiFi networks form a PPP of intensity $\lambda$ and each network is independently active with probability $p$. Hence, it is effectively a PPP of intensity $p\lambda$ with all nodes being active simultaneously. Furthermore, we consider a fixed transmit power for all WiFi networks (which is common in practice) of $P_x=P_{SU}$.

The aggregate interference $I_{aggr}$ in (\ref{ch5:eq:shotnoise}) is a weighted sum of received power from many independent APs that are distributed over a large area. Hence, it is reasonable to assume that $I_{aggr}$ has a Gaussian distribution with mean and variance of $\mu_I$ and $\sigma^2_I$, respectively\cite{ch5:paper:Sung11}. For a PPP of density $\lambda$, the mean and variance of the sum $\sum_{x\in\Phi}{f(x)}$ is calculated from Campbell's theorem \cite{ch5:Haenggi:2009:ILW} as
\begin{align}
\label{ch5:eq:avg-ppp}
E\left[ \sum_{x\in\Phi}{f(x)} \right] = \lambda \int_{\mathcal{R}^d}{f(x)dx}\\
var\left[ \sum_{x\in\Phi}{f(x)} \right] = \lambda \int_{\mathcal{R}^d}{f^2(x)dx}
\label{ch5:eq:var-ppp}
\end{align}

\subsection{Average and Variance of Interference ($\mu_I$, $\sigma_I^2$)}
The average interference that is received at radar receiver is calculated from (\ref{ch5:eq:avg-ppp}) by integrating over the $\mathcal{R}^2$ plane. Assuming that radar receiver is at the origin ($y=0$):
\begin{align}
\mu_I = E[I_{aggr}] = \frac{p\lambda  P_{SU}}{\mbox{FDR}(\Delta f)} \int_{\mathcal{R}^2}{G(\theta_x)l(||x||)dx}
\end{align}
Here we assume there is a minimum separation distance between radar and SU which could potentially be a function of $\theta$, $d(\theta)$. Using polar coordinates for the integral and considering $l(r)=K_0 r^{-\alpha}$ we obtain:
\ifStyleDouble
\begin{align}
\mu_I &= \frac{p\lambda  P_{SU} K_0}{\mbox{FDR}(\Delta f)} \int_{\theta}{\int_{d(\theta)}^{\infty}{G(\theta)r^{1-\alpha} dr d\theta}} \nonumber \\
      &= C_{\mu_I} \int_{\theta}{G(\theta) d^{2-\alpha}(\theta) d\theta} \nonumber \\
C_{\mu_I} &= \frac{p\lambda  P_{SU} K_0}{\mbox{FDR}(\Delta f)(\alpha - 2)}
\label{ch5:eq:Iaggr-mean}
\end{align}
\else
\begin{align}
\mu_I &= \frac{p\lambda  P_{SU} K_0}{\mbox{FDR}(\Delta f)} \int_{\theta}{\int_{d(\theta)}^{\infty}{G(\theta)r^{1-\alpha} dr d\theta}} 
      = C_{\mu_I} \int_{\theta}{G(\theta) d^{2-\alpha}(\theta) d\theta} \nonumber \\
C_{\mu_I} &= \frac{p\lambda  P_{SU} K_0}{\mbox{FDR}(\Delta f)(\alpha - 2)}
\label{ch5:eq:Iaggr-mean}
\end{align}
\fi
whenever $\alpha > 2$ is necessary to guarantee convergence of inner integral. This excludes the ideal `free space' ($\alpha = 2$) but holds for all practical scenarios of interest.

This equation allows variation in protection distance according to current direction of the radar's main antenna beam. The total interference highly depends on the choice of function $d(\theta)$ as a systematic parameter that trade-offs protection distances between main beam interferer versus side lobe ones. This is clearly chosen based on $G(\theta)$ and optimized subject to some constraints, as shown in the following sections.

The variance of aggregated interference is calculated from (\ref{ch5:eq:var-ppp}). Similar to the approach taken for $\mu_I$, the variance $\sigma_I^2$ is calculated by the following double integral over $r$ and $\theta$:
\ifStyleDouble
\begin{align}
\sigma_I^2 &= \frac{p\lambda P_{SU}^2 K_0^2}{\mbox{FDR}^2(\Delta f)} \int_{\theta}\int_{d(\theta)}^{\infty}{G^2(\theta) r^{-\alpha} rdrd\theta} \nonumber \\ 
           &= C_{\sigma_I^2} \int_{\theta}{G^2(\theta)d^{2-2\alpha}(\theta)d\theta} \nonumber \\
C_{\sigma_I^2} &= \frac{p\lambda P_{SU}^2 K_0^2}{\mbox{FDR}^2(\Delta f)(2\alpha-2)}
\label{ch5:eq:Iaggr-var}
\end{align}
\else
\begin{align}
\sigma_I^2 &= \frac{p\lambda P_{SU}^2 K_0^2}{\mbox{FDR}^2(\Delta f)} \int_{\theta}\int_{d(\theta)}^{\infty}{G^2(\theta) r^{-\alpha} rdrd\theta} = C_{\sigma_I^2} \int_{\theta}{G^2(\theta)d^{2-2\alpha}(\theta)d\theta} \nonumber \\
C_{\sigma_I^2} &= \frac{p\lambda P_{SU}^2 K_0^2}{\mbox{FDR}^2(\Delta f)(2\alpha-2)}
\label{ch5:eq:Iaggr-var}
\end{align}
\fi
with the assumption that $\alpha > 1$ to ensure convergence of the integration over $r$.

\subsection{Protection Region}
With multiple secondary users being active simultaneously, protection region for radar can be defined in terms of {\em probability of outage}, i.e. probability of effective radar SINR dropping below the minimum threshold. This is also equivalent to limiting aggregate interference $I_{aggr} < I_{\max}$. Since $I_{aggr}$ has a normal distribution $N(\mu_I, \sigma^2_I)$, the outage probability can be determined as following:
\begin{align}
P_{outage} = \mbox{Pr}\{ I_{aggr} > I_{\max} \} 
           = Q\left( \frac{I_{\max} - \mu_I}{\sigma_I} \right)
\end{align}
where $Q(.)$ function is the tail probability of the standard normal distribution. It is desired to set an upper bound for probability of outage, $P_{out,\max}$:
\begin{align}
& P_{outage} \le P_{out,\max} \rightarrow Q\left( \frac{I_{\max} - \mu_I}{\sigma_I} \right) \le  P_{out,\max} \nonumber \\
& I_{\max} \ge \mu_I + \sigma_I Q^{-1}\left( P_{out,\max} \right)
\label{ch5:eq:disrb-constraint}
\end{align}
This equation defines the relationship between maximum tolerable interference by the radar receiver and average/variance of aggregate interference from secondary WiFi networks. Depending on how much information about radar rotation is available at the SU, different scenarios are plausible for determining protection distance $d(\theta)$. Here we consider three special cases. \\
{\em First}: A secondary network that has {\em full knowledge} about current radar antenna beam position with respect to its location. \\
{\em Second}, the case of a SU that has {\em no knowledge} about radar rotation pattern and therefore is not capable of synchronizing its transmission instances with it. This results in a constant $d(\theta) = d_{\min}$ and a circular protection region.  \\
{\em Third}, the case of secondary user that is {\em partially aware} of radar's rotation schedule and is capable of identifying radar's main lobe from side lobe (representing a pragmatic, intermediate scenario between the above two).

Estimates of the protection distance is sensitive to the choice of the path loss model adopted. We use the well-known Longley-Rice (L-R) model that is based on field measurements and is relatively more accurate. However, previous analysis needs a closed form attenuation function of the form $l(r)=K_0 r^{-\alpha}$. Accordingly, we performed exponential curve fitting on L-R with parameters in tables \ref{ch5:tb:SU-param} and \ref{ch5:tb:rdr-aeronautical-B} to estimate $\alpha$ and $K_0$. L-R defines three propagation regions, namely {\em line of sight}, {\em diffraction} and {\em scattering}. By using line-of-sight region for curve fitting, $l(r)=259\,r^{-3.97}$ is obtained.

\subsubsection{Optimal Distance}
Using (\ref{ch5:eq:disrb-constraint}), the coexistence criteria is defined by limiting average and variance of aggregate interference $\mu_I + \sigma_I Q^{-1}\left( P_{out,\max} \right) \leq I_{\max}$. This inequality has a trivial answer that is achieved by letting $d(\theta)\rightarrow\infty$ (apparent from (\ref{ch5:eq:Iaggr-mean}) for example). In order to avoid this, we minimize the total protection area subject to net interference limit as formulated in following optimization problem:
\begin{equation}
d_{opt} = \mbox{arg}\min_{d(\theta)} \int_0^{2\pi}{\frac{d^2(\theta)}{2}d\theta}
\label{ch5:eq:opt_area}
\end{equation}
subject to:
\begin{equation}
\mu_I + \sigma_I Q^{-1}\left( P_{out,\max} \right) \leq I_{\max}
\label{ch5:eq:opt_constraint}
\end{equation}
For the most general antenna pattern model of $G(\theta)$, it is proven in the appendix that optimum protection distance $d_{opt}(\theta)$ is proportional to $G^{1/\alpha}(\theta)$ with a constant that is determined by numerically solving following equation:
\begin{align}
&d_{opt}(\theta) = \gamma G^{\frac{1}{\alpha}}(\theta) \nonumber \\
&\mathcal{A}\gamma^{2-\alpha} + \mathcal{B}\gamma^{1-\alpha} - I_{\max} = 0
\label{ch5:eq:d_opt_Multiple}
\end{align}
in which $\mathcal{A}$ and $\mathcal{B}$ are determined by:
\begin{align}
&\mathcal{A} = C_{\mu_I} \int_{0}^{2\pi}{G^{\frac{2}{\alpha}}(\theta)d\theta} \nonumber \\
&\mathcal{B} = Q^{-1}\left( P_{out,\max} \right) \sqrt{ C_{\sigma^2_I} \int_{0}^{2\pi}{G^{\frac{2}{\alpha}}(\theta)d\theta}  }
\end{align}

\subsubsection{Radar-Blind SU} 
For this type of SU, protection distance $d(\theta) = d_{\min}$ is constant and it simplifies equations (\ref{ch5:eq:Iaggr-mean}) and (\ref{ch5:eq:Iaggr-var}). By using simplified mean and variance in (\ref{ch5:eq:opt_constraint}), $d_{\min}$ is found as the solution of following equation:
\ifStyleDouble
\begin{align}
&d_{\min}^{2-\alpha} \left[ C_{\mu_I}\int{G(\theta)d\theta} \right] + \nonumber \\ 
&d_{\min}^{1-\alpha} \left[ Q^{-1}(P_{out,\max}) \sqrt{C_{\sigma_I^2} \int{G^2(\theta)d\theta}} \right] = I_{\max}
\end{align}
\else
\begin{align}
d_{\min}^{2-\alpha} \left[ C_{\mu_I}\int{G(\theta)d\theta} \right] + d_{\min}^{1-\alpha} \left[ Q^{-1}(P_{out,\max}) \sqrt{C_{\sigma_I^2} \int{G^2(\theta)d\theta}} \right] = I_{\max}
\end{align}
\fi

\subsubsection{Main/Side Lobe Interferer} 
While Equation (\ref{ch5:eq:d_opt_Multiple}) determines best protection distance in its general form, SUs have limited resolution in synchronizing with radar rotation in any practical scenario. A more pragmatic assumption is that secondaries can estimate when radar's main antenna beam is directed toward their location and stop their transmission accordingly. Here, radar antenna pattern is approximated as having two regions - a constant gain {\em main lobe} with a width of $\theta_H$ and constant gain {\em side lobe} that is $2\pi-\theta_H$ wide. Protection distance is similarly two distances - $d_{\max}$ and $d_{\min}$ for the main lobe and side lobe, respectively. Specifying one of these two distances allows the other to be calculated from total interference constraint. This degree of freedom allows us to optimize (minimize) total protection distance.

Let $\beta=\frac{d_{\max}}{d_{\min}}$ be the radio of main beam protection distance to side beam. For any choice of $\beta$, protection distances $d_{\min}$ and $d_{\max}$ can be determined from the constraint in (\ref{ch5:eq:opt_constraint}), which results in following:
\begin{align}
& d_{\min}^{2-\alpha} C_{\mu_I}\xi_1 + d_{\min}^{1-\alpha} Q^{-1}(P_{out,\max}) \sqrt{C_{\sigma_I^2} \xi_2} = I_{\max}   \nonumber \\
& \xi_1 = \int_{\frac{\theta_H}{2}}^{2\pi-\frac{\theta_H}{2}}{G(\theta)d\theta} + \beta^{2-\alpha} \int_{\frac{-\theta_H}{2}}^{\frac{\theta_H}{2}}{G(\theta)d\theta}    \nonumber \\
& \xi_2 = \int_{\frac{\theta_H}{2}}^{2\pi-\frac{\theta_H}{2}}{G^2(\theta)d\theta} + \beta^{2-2\alpha} \int_{\frac{-\theta_H}{2}}^{\frac{\theta_H}{2}}{G^2(\theta)d\theta} \nonumber \\
& d_{\max} = \beta d_{\min}
\end{align}

\begin{figure}[!t]%
	\centering
	\includegraphics[width=\figwidth]{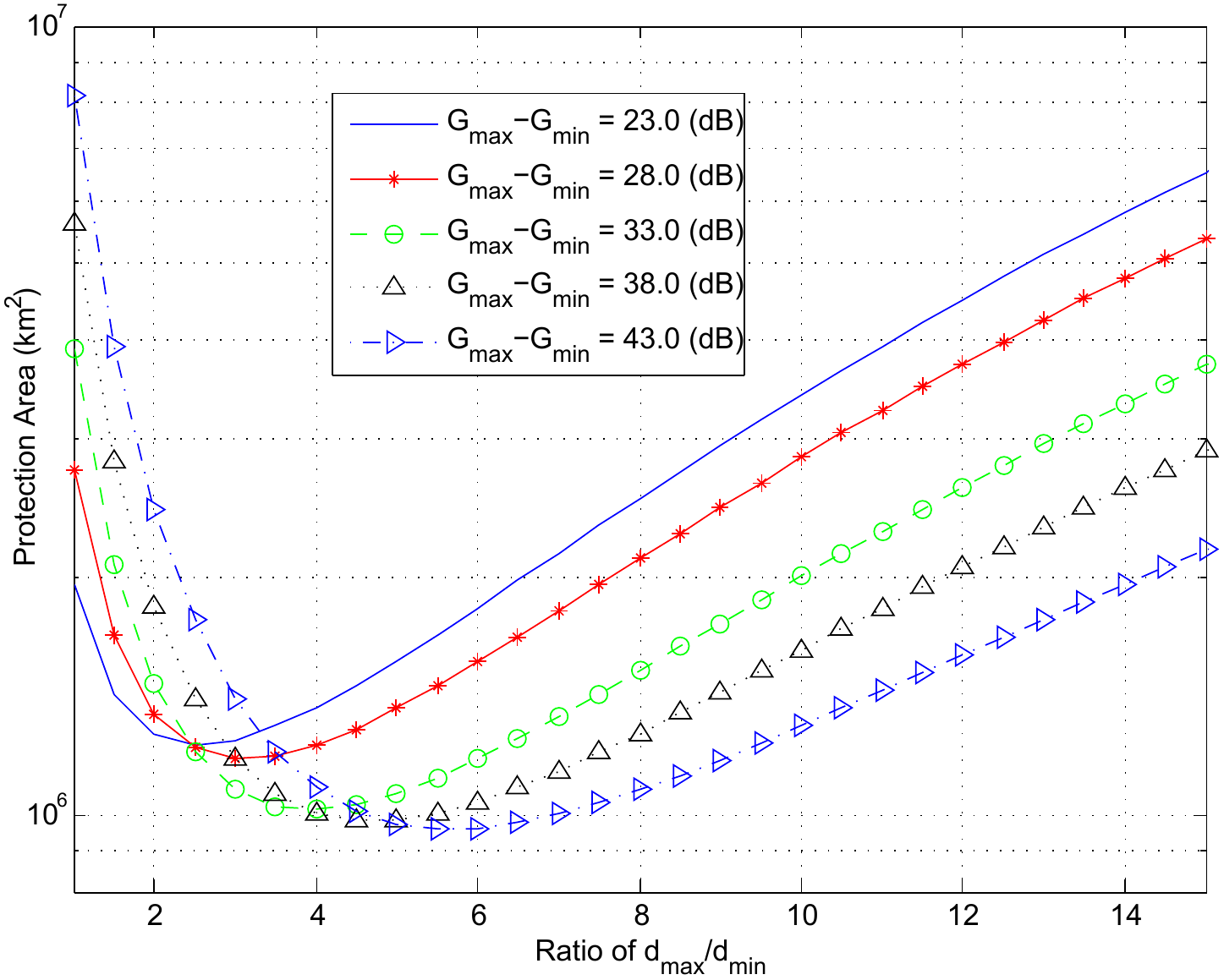}
	\caption{Total protection area v.s. $\frac{d_{\max}}{d_{\min}}$ ratio for various values of $\frac{G_{\max}}{G_{\min}}$; $p\lambda = 10^{-6}$.}
	\label{ch5:fig:prt-area-multi}
\end{figure}

The best ratio $\beta$ is selected to minimize total protection area of $Area = [\beta^2\theta_H/2 + \pi-\theta_H/2]d_{\min}^2$. For $p\lambda = 10^{-6}$ (time-space density product), figure \ref{ch5:fig:prt-area-multi} shows total protection area as a function of $\beta$ for various values of $\frac{G_{\max}}{G_{\min}}$ (radio of maximum to minimum radar antenna gain).

\begin{figure}[!t]%
	\centering
	\includegraphics[width=\figwidth]{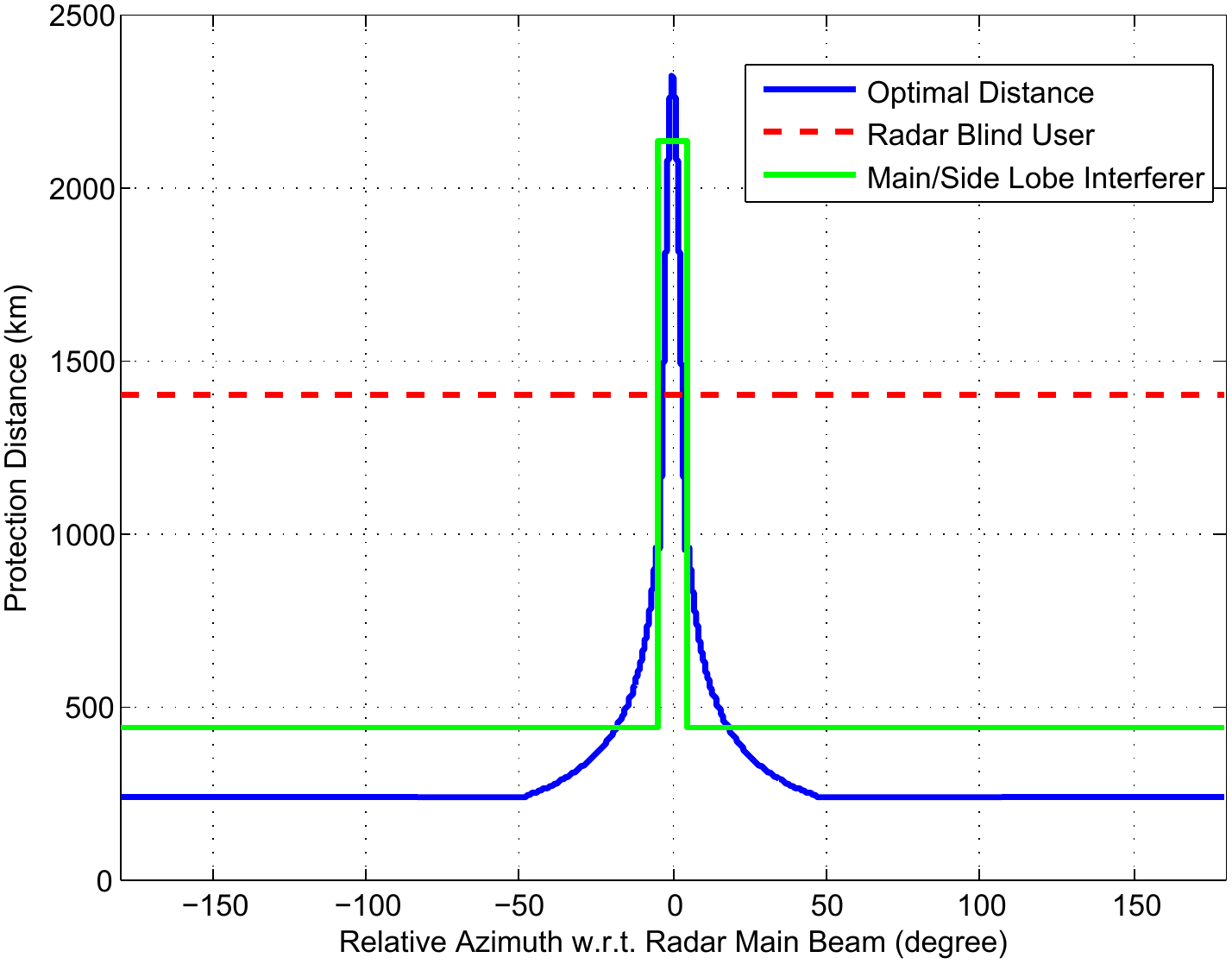}
	\caption{Protection region v.s. relative azimuth between SU and radar's main antenna beam for following cases: Radar-Blind SU, Optimal Protection Distance and Main/Side lobe interferer; $p\lambda = 10^{-6}$, $P_{out,\max}=0.1$}
	\label{ch5:fig:prt-rgn-multi-su}
\end{figure}

Figure \ref{ch5:fig:prt-rgn-multi-su} shows protection distance as a function of relative azimuth with radar's main beam for three cases of {\it Radar-Blind SU}, {\it Optimal Distance} and {\it Main/Side lobe interferer}. It is evident from this figure that a radar-blind SU will lose a significant portion of available white space spectrum as protection distance is significantly larger than other two cases. Main/Side interferer is plotted for the optimum choice of $\beta$. It provides a much closer distance to optimal results. A comparison between optimal distances here with that of single user in Figure \ref{ch5:fig:radarGainPrtRgn} reveals that distances are significantly increased (10-km for side lobes is expanded to 239-km) because of accumulated interference from spatial distribution of users.
Table \ref{ch5:tb:SU-area} compares total protection area for the three cases above. While total area occupied by {\it main/side lobe interferer} is about twice the optimal area, the required area for {radar-blind} user is 11.5 times larger than optimal, which again highlights the price to be paid for lack of information.

\begin{table}[ht!]
\caption{Protection Area Comparison}
\label{ch5:tb:SU-area}
\centering
\begin{tabular}{|c|c|c|c|}
\hline
\hline
 & {\bf Optimal} & {\bf Main/Side Lobe} & {\bf Radar Blind}\\
\hline
Total Area & 0.54 & 0.98 & 6.2 \\
(1000,000 $km^2$)&   &  & \\
\hline
Min. Distance (km) & 239 & 437 & 1403\\
\hline
Max. Distance (km) & 2331 & 2140 & 1403\\
\hline
\end{tabular}
\end{table}

\section{Interference to WiFi devices}
The main goal of spectrum sharing is to create new secondary networks, while providing protection to the incumbents (primary). Therefore, it is essential to study {\em primary to secondary} interference. In most sharing scenarios, secondary transmitters use a significantly lower power profile compared to the primary\footnote{For example, in TV white spaces, TV station may output up to 1000 KW as against portable secondary devices transmitting at 100 mW.}, rendering them very sensitive to interference from the primary.

The radar signal received at a WiFi receiver is given by
\begin{equation*}
y(t) = \sum_{n}\sqrt{\frac{G_{SU} G(\theta(t))P_T}{L_{Radar\rightarrow SU}}}s(t-\frac{n}{f_R})
\end{equation*}
where $G(\theta(t))$ is the instantaneous radar antenna gain and $L_{Radar\rightarrow SU}$ is the path loss from radar to SU. By reciprocity, the path loss from SU to radar is thus $L_{Radar\rightarrow SU}=\frac{1}{K_0r^{-\alpha}}$. The instantaneous interference power from the radar and resulting SINR at the input to the WiFi receiver can be written as
\begin{align}
&P_R(t) = P_T G_{SU} G(\theta(t)) K_0 d_{Radar-SU}^{-\alpha} \sum_n \Pi(\frac{t}{PW} - \frac{n}{f_R}) \nonumber \\
&\mbox{SINR}_{SU}(t) = \frac{P_{SU} G_{SU}}{L_{SU-SU} (N_0BW + P_R(t))}
\label{ch5:eq:su-sinr}
\end{align}

\begin{figure}[!t]%
	\centering
	\includegraphics[width=\figwidth]{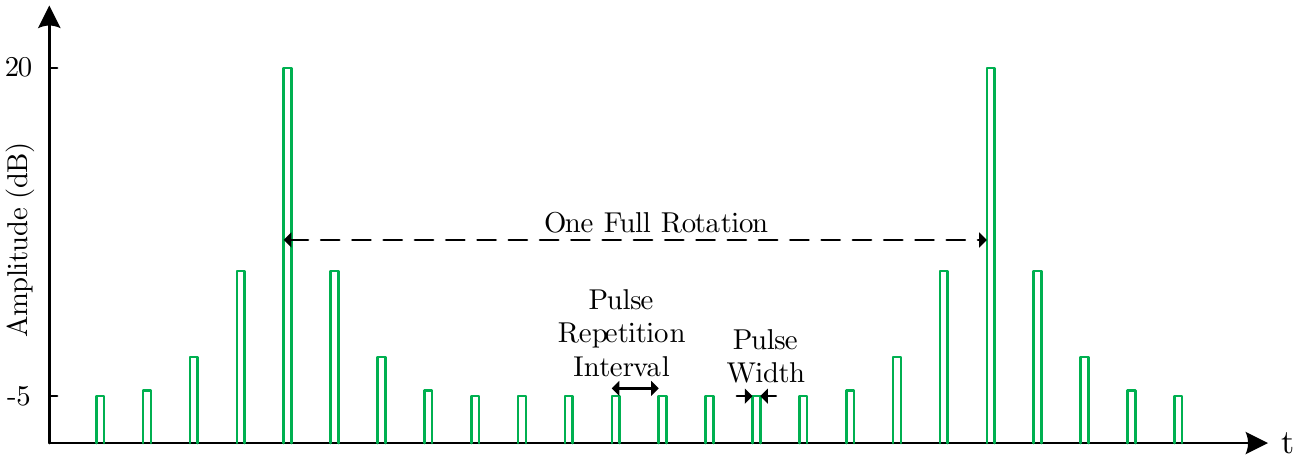}
	\caption{Radar signals received at WiFi receiver behave as a non-stationary source of interference.}
	\label{ch5:fig:radar-pulses}
\end{figure}

The radar interference to WiFi receivers is {\em non-stationary} for two reasons. {\em First}, due to radar rotation, the interference power varies periodically as a characteristic for search radars. Depending on rotation speed, this period is typically of the order of seconds. {\em Second}, the transmitted signals by radar $s(t-n/f_R)$ consists of short pulses as shown in Figure \ref{ch5:fig:radar-pulses}. For our typical aeronautical radar, the pulse width is 1$\mu s$ and pulse repetition internal is about 1$ms$. Therefore, even when the radar main beam is directly aligned with WiFi receiver ($G(\theta(t))$ is maximum), there are inter-pulse durations with zero interference.

Analytical evaluation of WiFi performance against a non-stationary interferer such as a pulsed radar is substantially more complicated than stationary ones for several reasons. {\em First}, depending on WiFi packet size and radar pulse repetition interval, the impact of radar signal on WiFi packet reception can vary greatly. For example, packet lengths in 802.11n can be vary from few hundreds of microseconds to several tens of milliseconds. Therefore, for pulse repetition interval of 1$ms$, short packets can fall in between inter-pulse intervals with significant probability, while longer packets almost surely overlap with radar pulses. {\em Second}, WiFi packets are composed of multiple OFDM symbols each of duration 4$\mu s$ \cite{ch5:paper:802.11n}. A radar pulse of 1-$\mu s$ width will collide with one symbol (or few symbols when packet is very long) out of many in the packet. Depending on the channel code (convolutional or LDPC) and selected MCS as well as the SNR of the interference-free channel, packet might still be decodeable. In addition, certain OFDM symbols are more crucial than the others. A collision between PLCP header and radar pulses will leave the entire packet undecodeable while impacted data symbols may be recovered by interleaving and channel coding. {\em Third}, all practical implementations of WiFi MAC/PHY layers include rate adaptation mechanisms to choose the best MCS based on channel condition. These algorithms are typically designed to converge to a steady state response in presence of {\em stationary} noise and interference. A non-stationary interferer can degrade the performance drastically unless smarter adaptation methods are designed which are aware of coexistence scenario.

A comprehensive WiFi performance study that considers all the aforementioned concerns is beyond the scope of this paper. Here, our focus is the achievable throughput in WiFi given the sharing scenario. Therefore, we assume that rate adaptation mechanism in WiFi always selects the best MCS for the current SINR. We consider a pair of 802.11n-based SUs in a 20-MHz channel with one spatial stream (1x1 SISO). The standard modulation and coding schemes in 802.11n as well as achievable rates are shown in Table \ref{ch5:tb:mcs}. The minimum required SNR for each MCS, corresponding to a 10\% packet loss, is also provided. The SNR values are obtained from \cite{ch5:thesis:daniel-uw} which are based on experimental measurements on an Intel Wireless Wi-Fi Link 5300 a/g/n.

\begin{table}[!ht]
\caption{Standard modulation and coding schemes and achievable data rates for 802.11n specifications. Minimum required SNR for each MCS, corresponding to 10\% packet loss, is also provided.}
\label{ch5:tb:mcs}
\centering
\begin{tabular}{ccccc}
\hline
\hline
{\bf MCS} & {\bf Modulation} & {\bf Coding Rate} & {\bf Data Rate(Mbps)} & {\bf SNR}\\
\hline
0 & BPSK   & 1/2 & 6.5  & 4.5\\
1 & QPSK   & 1/2 & 13.0 & 6.5\\
2 & QPSK   & 3/4 & 19.5 & 8.0\\
3 & 16-QAM & 1/2 & 26.0 & 10.5\\
4 & 16-QAM & 3/4 & 39.0 & 13.5\\
5 & 64-QAM & 2/3 & 52.0 & 17.5\\
6 & 64-QAM & 3/4 & 58.5 & 19.5\\
7 & 64-QAM & 5/6 & 65.0 & 21.5\\
\hline
\end{tabular}
\end{table}

The achievable throughput $R_{SU}(t)$ is a function of two factors; the instant SINR as in (\ref{ch5:eq:su-sinr}) that determines date rate through Table \ref{ch5:tb:mcs} and the fraction of time SU is allowed to transmit, $\rho(d)$. From previous analysis for a single user or multiple users, there is a minimum separate distance $d(\theta(t))$ that depends on the direction of radar's main beam. $\rho(d)$ defines the fraction of time when $d_{Radar-SU} \geq d(\theta(t))$. Therefore, SU throughput is:
\begin{align}
R_{SU}(t) = \left\{
      \begin{array}{lr}
				f(\mbox{SINR}(t)) &  d_{Radar-SU} \geq d(\theta(t))\\
				0 &  d_{Radar-SU} < d(\theta(t))\\
			\end{array}
            \right.
\end{align}
Therefore, a closer SU to radar has a lower throughput not only due to reduced SINR but also diminished transmission opportunity. The $\rho(d)$ factor also depends on our sharing policy. For example, for a {\em Radar-Blind} SU, $\rho(d)$ is a binary function while for {\em Main/Side lobe interferer}, it is constant as long as $d<d_{\max}$.

\begin{figure}[!ht]%
	\centering
	\includegraphics[width=\figwidth]{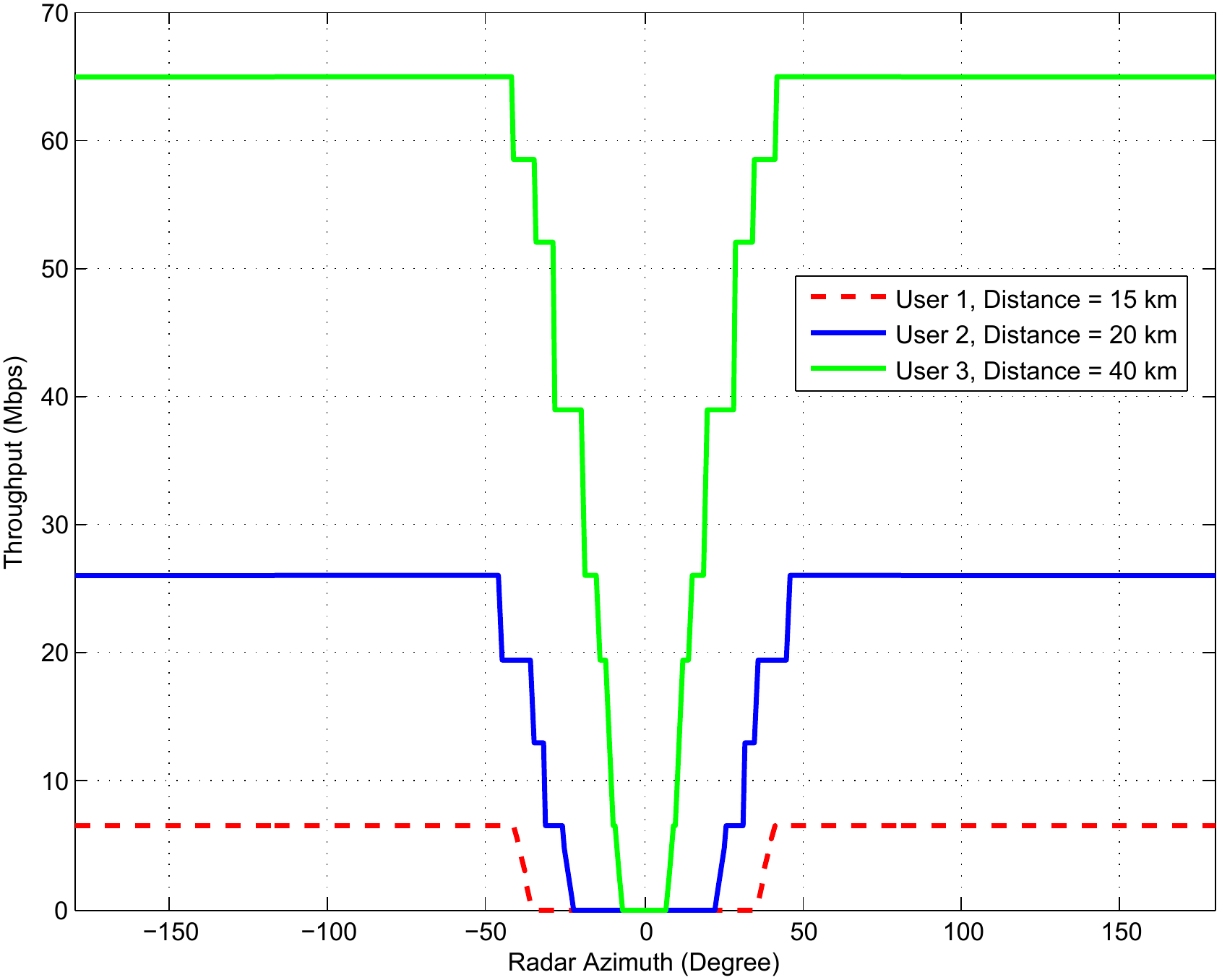}
	\caption{Secondary user throughput for single-SU sharing with radar.}
	\label{ch5:fig:single-su-sharing}
\end{figure}

Figure \ref{ch5:fig:single-su-sharing} shows achievable throughput by SU for a single user sharing scenario with radar. Three different users are considered at different distances from the radar and throughput variation is depicted with respect to radar rotation. The path loss between WiFi AP and station is set to 80 dB, corresponding to free space loss for a 100-meter link at 2.7GHz. As a first order approximation, SINR is set to instantaneous SINR as defined by (\ref{ch5:eq:su-sinr}), treating radar pulses as a continuous waveform (CW) interfering with WiFi OFDM symbols.

\begin{figure}[!t]%
	\centering
	\includegraphics[width=\figwidth]{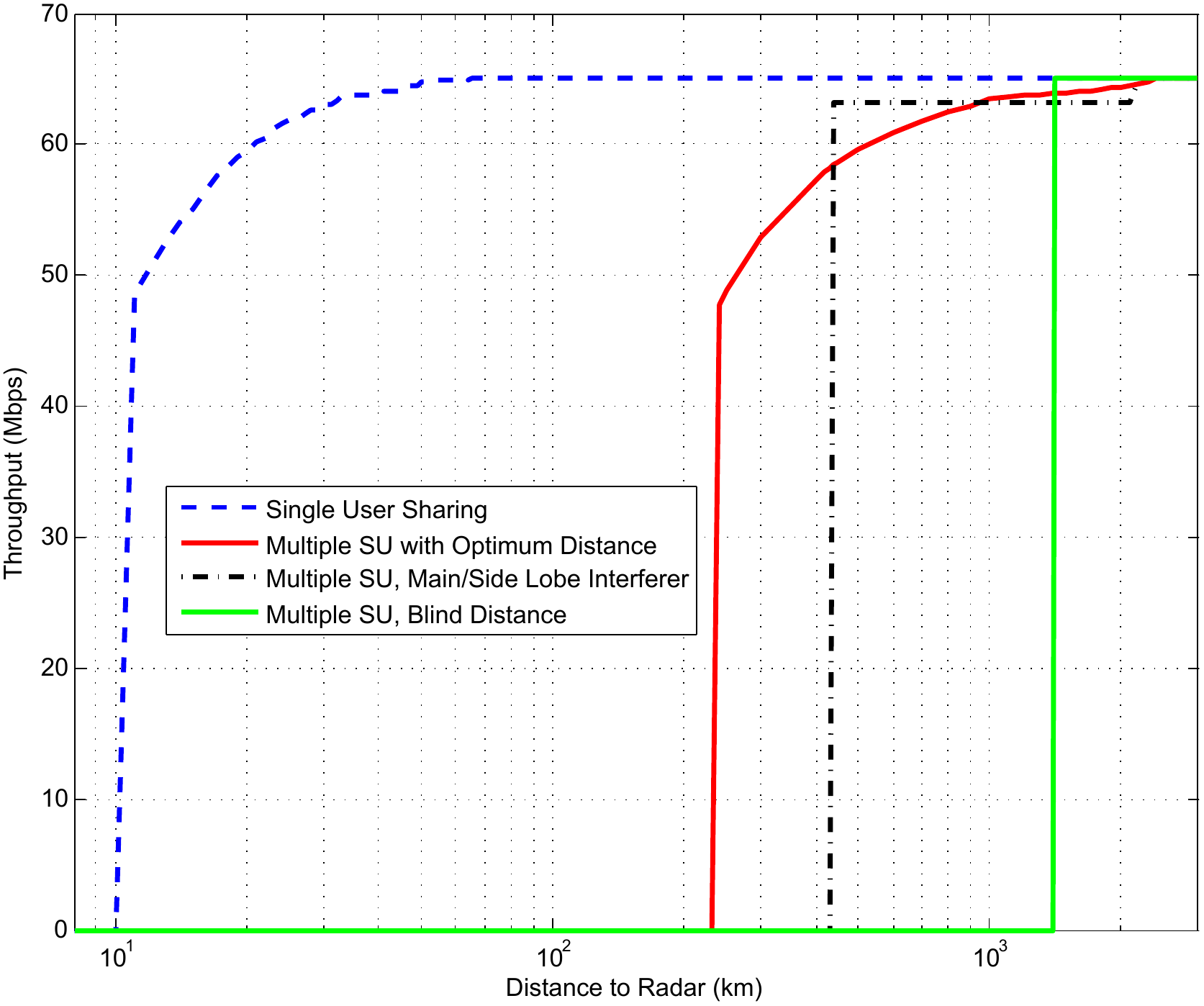}
	\caption{Average secondary user throughput for single/multiple SU sharing with radar.}
	\label{ch5:fig:msu-sharing}
\end{figure}

In order to differentiate radar pulses from a CW signal, we need to estimate effective SINR from (\ref{ch5:eq:su-sinr}). Since WiFi data are interleaved in time, an OFDM symbol (4-$\mu$s long) that falls within a radar pulse is later extended to (after de-interleaving) a significantly larger time interval. This is equivalent to extending radar pulse width while reducing its power level. Therefore, if we assume that WiFi interleaver is sufficiently long, effective radar interference is $\overline{P_R(t)} = \mbox{PW}f_R P_T G_{SU} G(\theta(t)) K_0 d_{Radar-SU}^{-\alpha}$, which is averaged over pulse repetition interval. Here, radar interference to WiFi receiver is scaled by a factor of pulse width/pulse repetition interval. The average SINR and throughput experienced by the SU is:
\begin{align}
\overline{\mbox{SINR}(t)}   &=  \frac{P_{SU} G_{SU}}{L_{SU-SU} (N_0BW + \overline{P_R(t)})} \nonumber \\
\overline{R_{SU}} &= \int_{<T_S>}{\overline{R_{SU}(t)}dt}
\label{ch5:eq:avg-throughput}
\end{align}
where integration is over the scan time of radar, $T_S$. Figure \ref{ch5:fig:msu-sharing} shows average SU throughput based on (\ref{ch5:eq:avg-throughput}) for single and multiple SU sharing.

\section{Numerical Results}
In this section, protection distance and SU throughput is evaluated against various systematic parameters. In previous sections, we made a conservative assumption that the SNR of radar pulses reflected from a target at the edge of radar's coverage area is already at the minimum required level specified by (\ref{ch5:eq:snr-pd-pfa}). This will leave only a small room for additional interference from secondaries and therefore results in a larger protection distance. In this work, we explore the impact of relaxing the SNR from (\ref{ch5:eq:SNR}) by allowing varying degrees of secondary interference.

\begin{figure}[!t]%
	\centering
	\includegraphics[width=\figwidth]{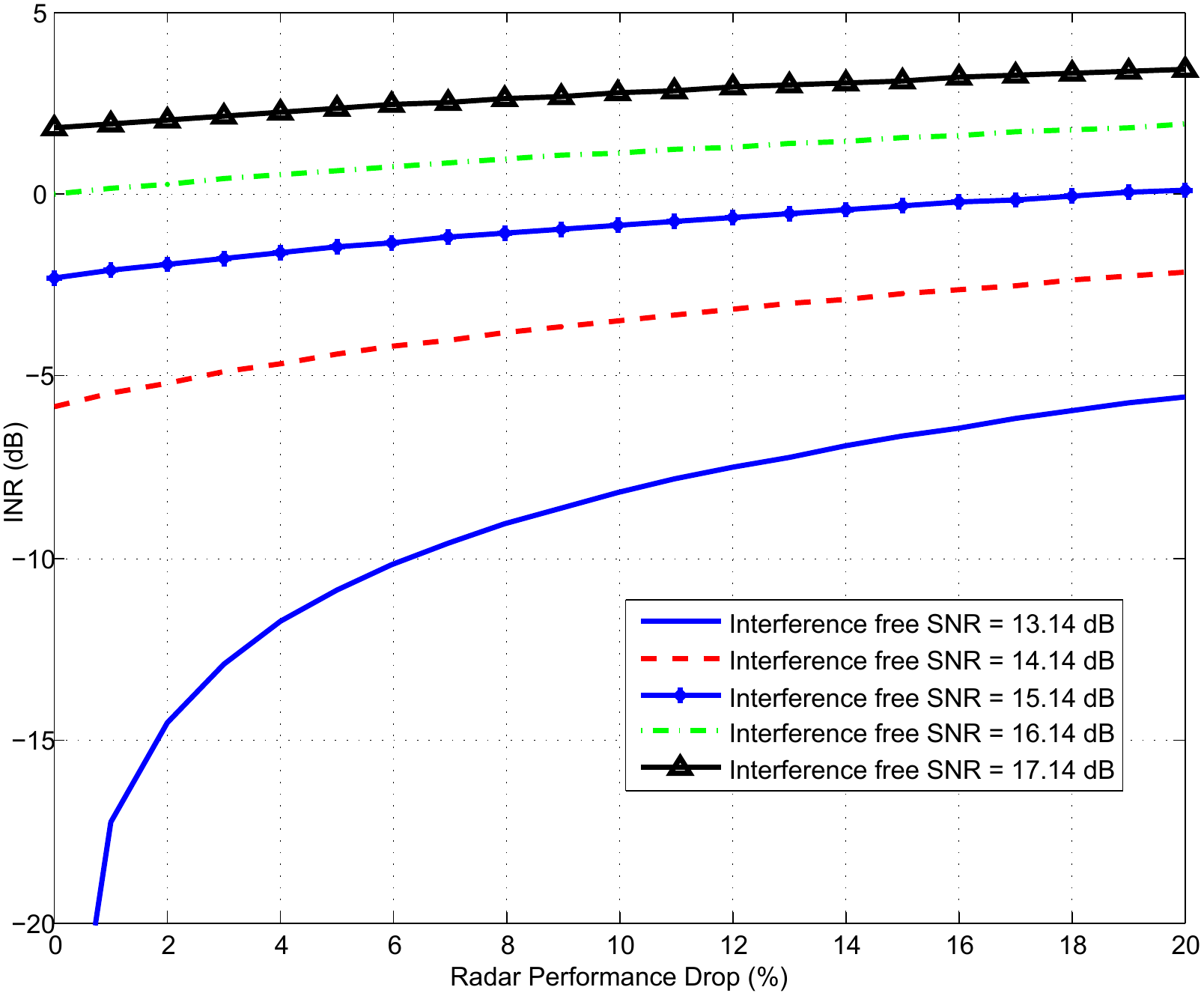}
	\caption{Maximum permitted INR versus radar performance drop for various values of original SNR (interference-free SNR)}
	\label{ch5:fig:inr-perf-drop}
\end{figure}

Figure \ref{ch5:fig:inr-perf-drop} shows maximum permitted INR caused by SU as a function of radar performance drop (reduction in $P_d$). Original $P_d$ is set to 0.9 and is allowed to drop to 0.7 for the results in this figure. Different curves correspond to various initial SNR (without SU interference) at the radar. Bigger initial SNRs open more room for external interference and result in larger INRs, as can be seen in this chart. For example at initial SNR of 16.14 and with 0\% performance drop, INR can be as large as 0 dB. This is because initial SNR is 3 dB above minimum required SNR of 13.14 dB, therefore a 3 dB drop is allowed.

Using parameters in Table \ref{ch5:tb:rdr-aeronautical-B}, the initial SNR at the radar input is estimated to be 30.57 dB which allows maximum INR of +17.69 dB for 5\% drop in $P_d$. We use this INR in the following to determine a less conservative protection distance then previous sections as well as achievable SU throughput.

\subsection{Protection Distance}
\begin{figure}[!t]%
	\centering
	\includegraphics[width=\figwidth]{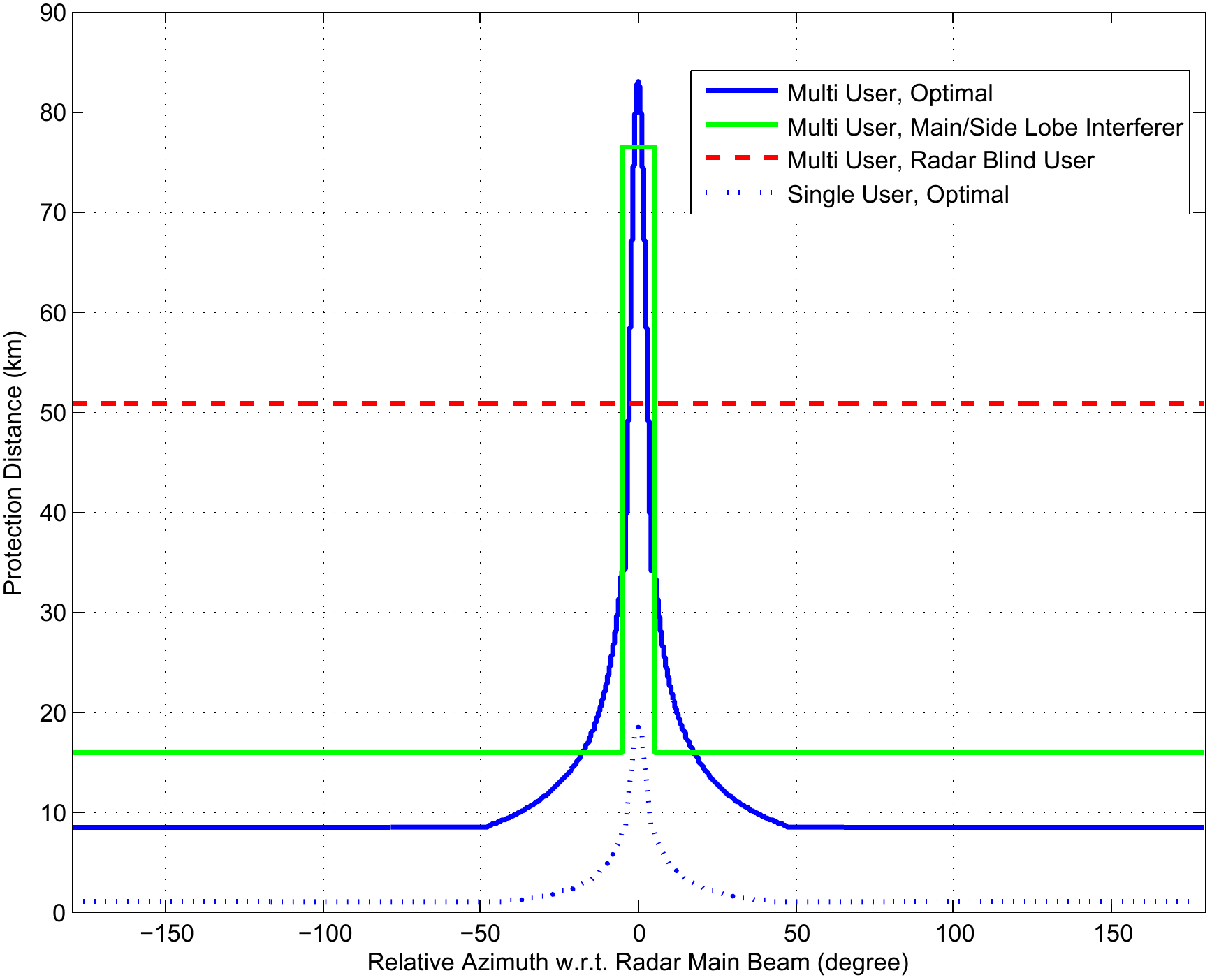}
	\caption{Protection distance versus azimuth for single and multi-user sharing}
	\label{ch5:fig:prtc-dist-less-conservative}
\end{figure}

\begin{figure}[!t]%
	\centering
	\includegraphics[width=\figwidth]{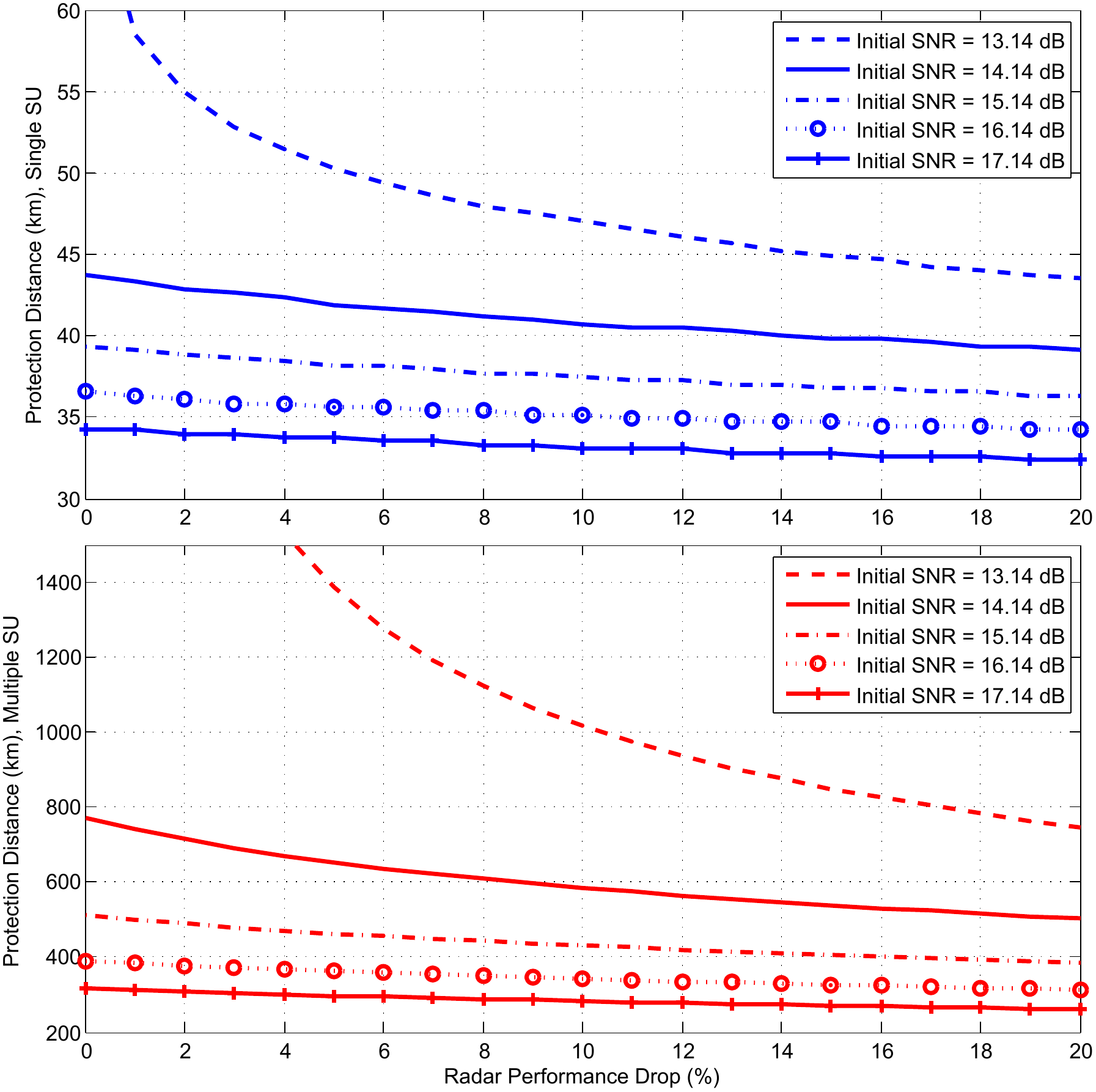}
	\caption{Protection distance versus radar performance drop for single/multiple radar-blind secondary users. Initial SNR corresponds to noise-limited SNR at radar receiver.}
	\label{ch5:fig:prtc-dist-vs-pddrop}
\end{figure}

Figure \ref{ch5:fig:prtc-dist-less-conservative} shows protection distances for single and multiple SU sharing. Comparing this with Figs. \ref{ch5:fig:radarGainPrtRgn} and \ref{ch5:fig:prt-rgn-multi-su} reveals that protection distances are immensely reduced because INR is increased from -10.96 dB to +17.69 dB. To better investigate the effect of initial radar SNR on required protection distances, Fig. \ref{ch5:fig:prtc-dist-vs-pddrop} shows protection distance for single/multiple radar-blind secondary users with different initial radar SNR. The minimum required SNR for target ROC point of $P_d$=0.90 and $P_{fa}=10^{-6}$ is 13.4 dB. Therefore, for the initial SNR of 13.14 dB, the allowed radar performance drop (because of additional interference) has significant impact on the protection distance. However, if initial SNR is above this limit by only a few dB, the dependency of protection distance on radar performance drop is significantly reduced. For example at SNR of 17.14 dB, increasing radar $P_d$ drop from 90\% to 70\% will reduce the distance from 315 km to 260 km (multiple SU, radar-blind).

\begin{figure}[!t]%
	\centering
	\includegraphics[width=\figwidth]{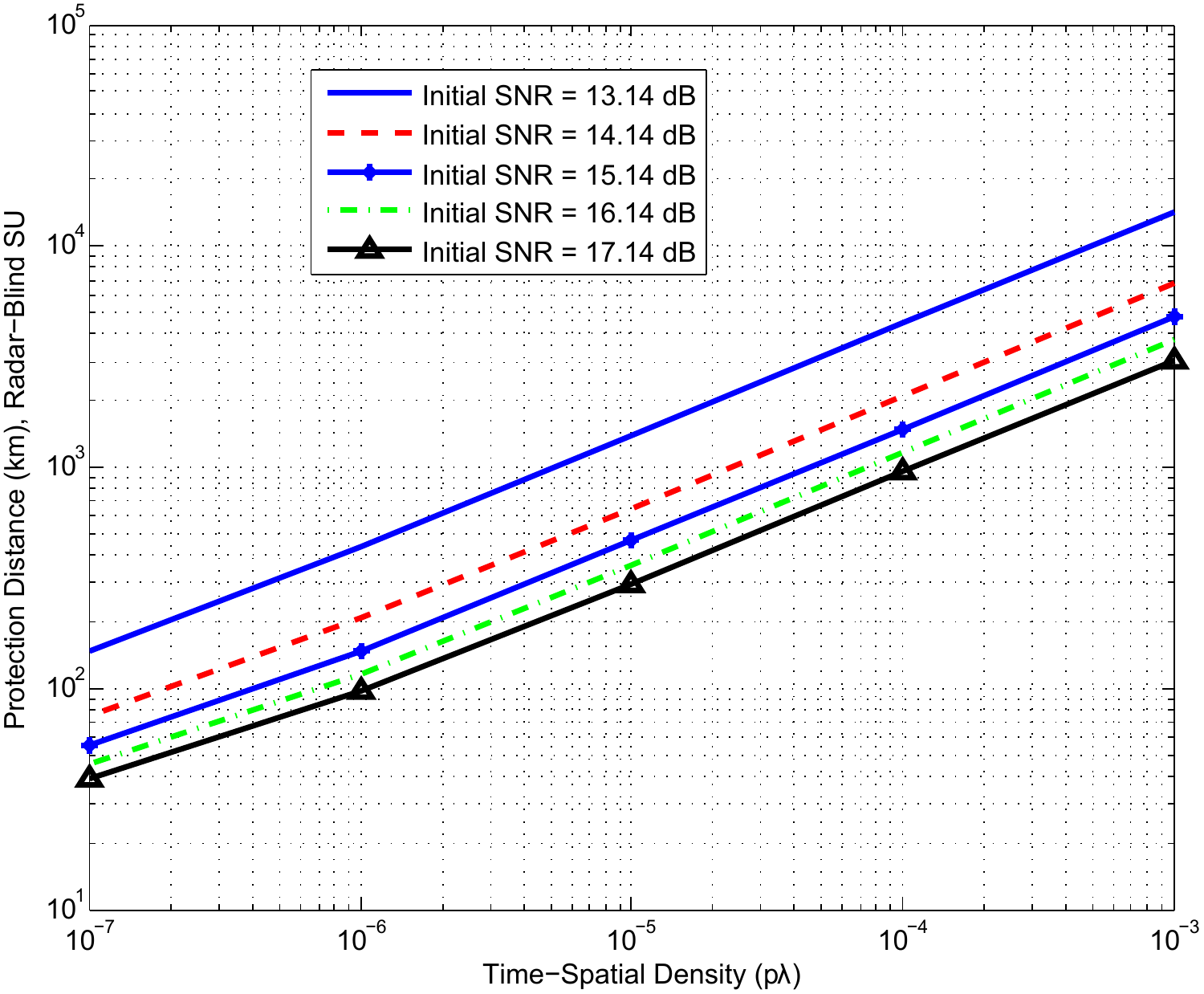}
	\caption{Protection distance versus time-spatial density of WiFi networks, $p\lambda$}
	\label{ch5:fig:prtc-dist-vs-plambda}
\end{figure}

In sharing radar spectrum with distributed SUs, the average interference is also highly affected by population density and probability of WiFi network's activity. Figure \ref{ch5:fig:prtc-dist-vs-plambda} shows this dependency by evaluating protection distance for radar-blind users versus the product of $p\lambda$ and for different initial radar SNR. A constant performance drop of 5\% is utilized for radar. It is clear from this figure that in the logarithmic scale, protection distance is a linear function of $p\lambda$.

\subsection{SU Throughput}
By increasing initial radar SNR or allowing further drop in its performance, we observed significant reductions in protection distances as shown in previous results. Reduced distances provide additional white space opportunities for WiFi devices. On the other hand, closer distances to radar means additional interference from transmitted pulses.

\begin{figure}[!t]%
	\centering
	\includegraphics[width=\figwidth]{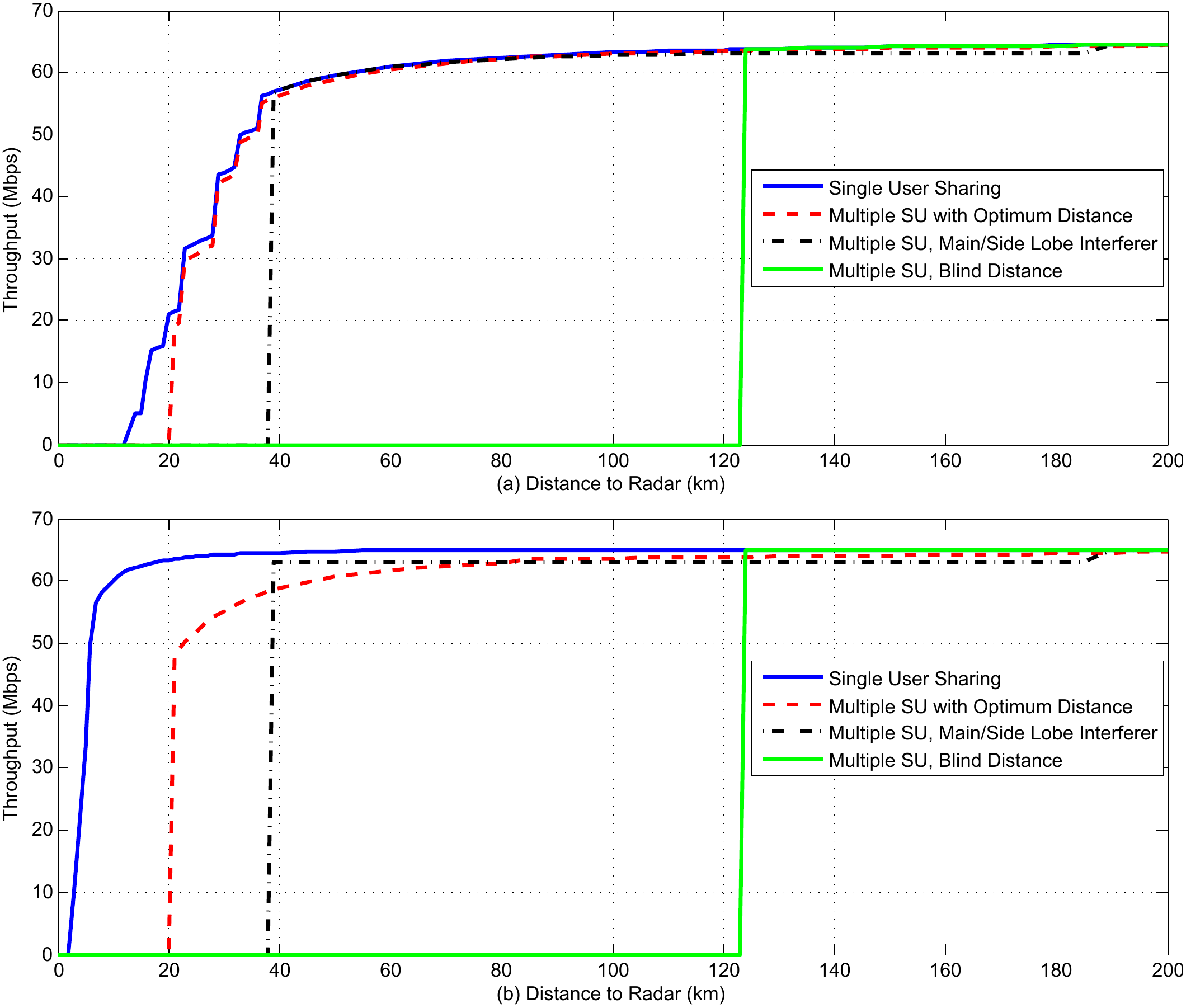}
	\caption{Achievable SU throughput versus distance for various sharing policies. (a) is based on peak radar interference to WiFi receiver and (b) is based on average radar interference.}
	\label{ch5:fig:throughput-dist-high-radar-snr}
\end{figure}

The average interference from radar to WiFi receiver was calculated in (\ref{ch5:eq:avg-throughput}) by scaling peak power with the ratio of pulse width to pulse repetition interval. For our radar parameters, this translates to $\frac{1 \mu s}{896 \mu s} \approx 29.5$ dB reduction in effective radar interference level which significantly improves WiFi SINR at close distances to radar. Figure \ref{ch5:fig:throughput-dist-high-radar-snr} shows achievable SU throughput for both cases of using peak radar interference (a) and average/effective radar interference (b) to WiFi receivers (29.5 dB reduction w.r.t. peak). Initial radar SNR is set to 23.14-dB which is 10-dB above minimum required level and radar performance drop is set to 5\%. For a radar-blind SU that can only coexist with radar at large distances of $>$120 km, throughput is the same in both cases because radar interference is negligible. However, at close distances of single-user sharing and multi-user with optimal distance, throughput drop due to radar interference is very clear in (a). Particularly for the case of single-user sharing, protection distance is reduced to about 2-km, but practical throughput is still zero up to 12 km from radar.

\section{Conclusion}
In this paper, we considered the problem of spectrum sharing between a rotating radar and WiFi networks. Minimum required SNR for noise-limited operation of the radar was defined as a function of basic radar parameters, including probability of {\em detection}. Coexistence with WiFi users was made possible by permitting a certain drop in radar's detection performance. We showed that this performance drop is very essential when radar SNR (without interference from WiFi users) is very close to the minimum required SNR. This determined maximum tolerable interference by the radar from WiFi devices (INR). Evaluating INR for various values of radar detection drops revealed that INR falls abruptly at small performance detection drops, when radar SNR is already at its minimum; otherwise INR changes are slow.

Protection distance - the minimum required distance between SU and radar receiver - was calculated for both single-SU case as well as multiple spatially distributed SUs. The latter formed a Poisson point process in space and an aggregate interference to radar that was approximated as Gaussian. Outage probability was utilized as the defining metric for protection distance calculation and different sharing scenarios was introduced based on how much radar-related data is available to the SU.

The optimal protection distance was defined in terms of minimizing total protected area. It was shown to be proportional to $G^{\frac{1}{\alpha}}(\theta)$. For a radar-blind SU, a constant protection distance was defined which was significantly larger than optimal distance. Comparing total protected area for these two showed that radar-blind area is about 12 times (for our settings) larger than optimal area. A more pragmatic solution is an SU with sufficient side information about radar to distinguish main lobe from side lobe. Protection distance for this type of SU was calculated and shown to be very close to optimal distance.

The effect of interference caused by radar pulses on performance of WiFi networks was modeled and achievable throughput (as a function of radar rotation as well as average) was estimated. For close distances to radar, throughput was shown to be very low even though SU is allowed to transmit. Since radar interference is non-stationary, two cases were considered as the upper and lower bounds of effective radar interference. First, instantaneous interference from radar pulses was utilized for calculating effective SINR. Second, the power of radar pulses was normalized by the ratio of pulse-width/pulse-repetition-interval. The former showed significant throughput reduction at close distance (single SU and optimal multiple SU).

\ifCLASSOPTIONcaptionsoff
  \newpage
\fi

\bibliographystyle{IEEEtran}
\bibliography{IEEEabrv,mylit}

\appendix
\subsection{Radar Parameters}{\label{ch5:app:radar}}
Radar parameters used for simulation purposes in this paper are presented in table \ref{ch5:tb:rdr-aeronautical-B}.
\begin{table}[ht!]
\caption{Technical Parameter for Type B Aeronautical Radar}
\label{ch5:tb:rdr-aeronautical-B}
\centering
\begin{tabular}{|c|c|}
\hline
\hline
{\bf Characteristics } & {\bf Radar B }\\
\hline
Platform Type & Ground, ATC \\
Tuning Range (MHz) &  2700 - 2900 \\
Modulation & P0N\footnotemark\\
Tx power into antenna & 1.32 MW\\
Pulse Width ($\mu s$)&   1.03 \\
Pulse rise/fall time ($\mu s$)&  -- \\
Pulse repetition rate (pps)&  1059 - 1172\\
Duty Cycle & 0.14 maximum \\
Chirp BW & NA \\
Compression Ratio & NA \\
RF emission BW (-20 dB) & 5 MHz \\
RF emission BW (3 dB) & 600 kHz \\
\hline
\textbf{Antenna Parameters} & \\
\hline
Type & Parabolic reflector \\
Pattern type (degrees) & Cosecant-squared +30 \\
Polarization & Vertical or right hand circular\\
Main beam gain (dBi) &  33.5  \\
Elevation beamwidth (degree) & 4.8 \\
Azimuthal beamwidth (degree) & 1.3 \\
Horizontal scan rate (degree/s) & 75 \\
Horizontal scan type\footnotemark (degrees) & 360 \\
Vertical scan rate (degree/s) & N/A \\
Vertical scan type (degree) & N/A \\
Side-lobe levels (1st and remote) & 7.3dBi \\
Height (m) & 8.0 \\
\hline
\textbf{Receiver Parameters} & \\
\hline
IF 3 dB bandwidth & 653 kHz \\
Noise figure (dB) & 4.0 maximum \\
Minimum discernible signal (dBm) & -108 \\
Receiver RF 3 dB bandwidth (MHz) & 10 \\
\hline
\end{tabular}
\end{table}
\setcounter{footnote}{2}
\footnotetext{P0N: No modulating signal and no information transmitted\cite{ch5:ntia:goverment-systems}}
\setcounter{footnote}{3}
\footnotetext{Options are: continuous, random, 360 $\deg$, sector, etc.}

\subsection{Optimum Protection Distance}
Based on equations (\ref{ch5:eq:opt_area}) and (\ref{ch5:eq:opt_constraint}), optimal protection distance by limiting maximum outage probability is obtained as:
\begin{align*}
&d_{opt} = \mbox{arg}\min_{d(\theta)} \int_0^{2\pi}{\frac{d^2(\theta)}{2}d\theta} \\
& \mu_I + \sigma_I Q^{-1}\left( P_{out,\max} \right) \leq I_{\max}
\end{align*}
where $\mu_I$ and $\sigma_I$ are calculated in (\ref{ch5:eq:Iaggr-mean}) and (\ref{ch5:eq:Iaggr-var}):
\begin{align*}
\mu_I &= C_{\mu_I} \int_{\theta}{G(\theta) d^{2-\alpha}(\theta) d\theta}  \\
\sigma_I &= \sqrt{ C_{\sigma_I^2} \int_{\theta}{G^2(\theta)d^{2-2\alpha}(\theta)d\theta} }
\end{align*}
Optimal $d(\theta)$ is attained by converting the inequality constraint to equality. This follows because for any $d(\theta)$ for which the strict inequality constraint holds, we can scale down $d(\theta)$ accordingly to increase $\mu_I$, $\sigma_I$ and achieve equality constraint (note that $2-\alpha<0$ and $2-2\alpha<0$). This will clearly result in a smaller objective function.

 With equality constraint, we use Lagrange multiplier method with a dummy variable $\epsilon$ to redefine objective function as
\ifStyleDouble
\begin{align*}
d_{opt} = \mbox{arg}\min_{d(\theta)} & \int_0^{2\pi}{\frac{d^2(\theta)}{2}d\theta} + \nonumber \\
                                     &\epsilon\left( \mu_I + \sigma_I Q^{-1}\left( P_{out,\max} \right) - I_{\max} \right)
\end{align*}
\else
\begin{equation*}
d_{opt} = \mbox{arg}\min_{d(\theta)} \int_0^{2\pi}{\frac{d^2(\theta)}{2}d\theta} + \epsilon\left( \mu_I + \sigma_I Q^{-1}\left( P_{out,\max} \right) - I_{\max} \right)
\end{equation*}
\fi
Taking partial derivatives of the new objective function with respect to $d(\theta)$ results in:
\begin{align*}
& \frac{\partial f}{\partial d(\theta)} = 0 \nonumber \\
& d(\theta) + \epsilon \left[  \frac{\partial\mu_I}{\partial d(\theta)} + Q^{-1}(P_{out,\max}) \frac{\partial\sigma_I}{\partial d(\theta)} \right] = 0
\end{align*}
Replacing $\mu_I$ and $\sigma_I$:
\ifStyleDouble
\begin{align*}
& d(\theta) + \epsilon \left[ (2-\alpha)C_{\mu_I}G(\theta)d^{1-\alpha}(\theta) + \right. \\
 &   \frac{Q^{-1}(P_{out,\max}) C_{\sigma_I^2}(2-2\alpha)G^2(\theta)d^{1-2\alpha}(\theta)}{2\sqrt{C_{\sigma_I^2} \int_{\theta}{G^2(\theta)d^{2-2\alpha}(\theta)d\theta}}}  \left. \right] = 0
\end{align*}
\else
\begin{align*}
& d(\theta) + \epsilon \left[ (2-\alpha)C_{\mu_I}G(\theta)d^{1-\alpha}(\theta) +
    \frac{Q^{-1}(P_{out,\max}) C_{\sigma_I^2}(2-2\alpha)G^2(\theta)d^{1-2\alpha}(\theta)}{2\sqrt{C_{\sigma_I^2} \int_{\theta}{G^2(\theta)d^{2-2\alpha}(\theta)d\theta}}}  \right] = 0
\end{align*}
\fi
Let $X=G(\theta)d^{-\alpha}(\theta)$, the above equation can be written as $1 + \epsilon \left[ \Gamma X + \Lambda X^2 \right] = 0$, where $\Lambda$ and $\Gamma$ are constant. Solving for $X$ results in $G(\theta)d^{-\alpha}(\theta)=\frac{-\epsilon\Gamma \pm \sqrt{\epsilon^2\Gamma^2-4\epsilon\Lambda}}{2\epsilon\Lambda}$. Therefore, $d(\theta)$ is proportional to $G^{\frac{1}{\alpha}}(\theta)$. The proportionality constant is found from the constraint equation:
\ifStyleDouble
\begin{align*}
& d(\theta) = \gamma G(\theta)^{\frac{1}{\alpha}} \\
& C_{\mu_I} \int_{\theta}{G(\theta) \gamma^{2-\alpha} G^{\frac{2-\alpha}{\alpha}}(\theta) d\theta} + \\
&Q^{-1}\left(P_{out,\max}\right)\sqrt{ C_{\sigma_I^2} \int_{\theta}{G^2(\theta)\gamma^{2-2\alpha}G^{\frac{2-2\alpha}{\alpha}}(\theta)d\theta} } = I_{\max}
\end{align*}
\else
\begin{align*}
& d(\theta) = \gamma G(\theta)^{\frac{1}{\alpha}} \\
& C_{\mu_I} \int_{\theta}{G(\theta) \gamma^{2-\alpha} G^{\frac{2-\alpha}{\alpha}}(\theta) d\theta} + Q^{-1}\left(P_{out,\max}\right)\sqrt{ C_{\sigma_I^2} \int_{\theta}{G^2(\theta)\gamma^{2-2\alpha}G^{\frac{2-2\alpha}{\alpha}}(\theta)d\theta} } = I_{\max}
\end{align*}
\fi
which is simplified to:
\ifStyleDouble
\begin{align*}
& \gamma^{2-\alpha} C_{\mu_I}\int_{\theta}{G^{\frac{2}{\alpha}}(\theta) d\theta} + \\
& \gamma^{1-\alpha} Q^{-1}\left(P_{out,\max}\right)\sqrt{ C_{\sigma_I^2} \int_{\theta}{G^{\frac{2}{\alpha}}(\theta)d\theta} } = I_{\max}
\end{align*}
\else
\begin{align*}
& \gamma^{2-\alpha} C_{\mu_I}\int_{\theta}{G^{\frac{2}{\alpha}}(\theta) d\theta} + \gamma^{1-\alpha} Q^{-1}\left(P_{out,\max}\right)\sqrt{ C_{\sigma_I^2} \int_{\theta}{G^{\frac{2}{\alpha}}(\theta)d\theta} } = I_{\max}
\end{align*}
\fi

\end{document}